\documentclass[11pt]{revtex4}
\usepackage{graphicx}
\usepackage{amsmath}
\usepackage{amssymb}

\begin{document}

\begin{title}{Do mirrors for gravitational waves exist?}\end{title}

\begin{center}
{\huge {Do Mirrors for Gravitational Waves Exist?} 
}\\ \vspace{0.1in} {\Large {Stephen Minter$^{a}$, Kirk Wegter-McNelly$^{b}$, and Raymond
Chiao$^{c}$}  }\\ \vspace{0.2in} \scriptsize{{$^{a}$University of
California, Merced, School of Natural Sciences, P.O. Box 2039, Merced, CA
95344, USA \\ \vspace{0.1in} $^{b}$Boston University, School of
Theology, 745 Commonwealth Ave., Boston, MA 02215, USA\\ \vspace{0.1in}
$^{c}$University of California, Merced, Schools of Natural Sciences and
Engineering, P.O. Box 2039, Merced, CA 95344, USA. Corresponding author,
email: rchiao@ucmerced.edu}}\\ \vspace{0.5in}

{\Large {\textbf{Abstract}} }
\end{center}

Thin superconducting films are predicted to be highly reflective mirrors for
gravitational waves at microwave frequencies. The quantum mechanical
\emph{non-localizability} of the negatively charged Cooper pairs, which is
protected from the localizing effect of decoherence by an energy gap, causes
the pairs to undergo \textit{non-picturable}, \textit{non-geodesic} motion in
the presence of a gravitational wave. This non-geodesic motion, which is
accelerated motion \emph{through }space, leads to the existence of mass and
charge supercurrents inside the superconducting film. On the other hand, the
decoherence-induced \emph{localizability} of the positively charged ions in
the lattice causes them to undergo \textit{picturable}, \textit{geodesic}
motion as they are carried along \textit{with} space in the presence of the same gravitational wave. The resulting
separation of charges leads to a virtual plasma excitation within the film
that enormously enhances its interaction with the wave, relative to that of a
neutral superfluid or any normal matter. The existence of strong mass
supercurrents within a superconducting film in the presence of a gravitational
wave, dubbed the \textquotedblleft Heisenberg-Coulomb
effect,\textquotedblright\ implies the specular reflection of a gravitational
microwave from a film whose thickness is much less than the London penetration
depth of the material, in close analogy with the electromagnetic case. The
argument is developed by allowing classical gravitational fields, which obey
Maxwell-like equations, to interact with quantum matter, which is described
using the BCS and Ginzburg-Landau theories of superconductivity, as well as a
collisionless plasma model. Several possible experimental tests of these
ideas, including mesoscopic ones, are presented alongside comments on the
broader theoretical implications of the central hypothesis.\\ 

\hspace{-0.21in}\textbf{Keywords}: gravitational wave, mirror, superconductor, uncertainty principle, equivalence principle, Heisenberg-Coulomb effect 

\section{Introduction}

Experiments at the frontiers of quantum mechanics and gravitation are rare. In
this paper we argue for a claim that may lead to several new types of
experiment, namely, that a superconducting film whose thickness is less than
the London penetration depth of the material can specularly reflect not only
electromagnetic (EM) microwaves, as has been experimentally demonstrated
\cite{Glover-and-Tinkham,Tinkham}, but gravitational (GR) microwaves as well.
The basic motivation for our approach lies in the well-known fact that
Einstein's field equations lead, in the limits of \emph{weak }GR fields and
\emph{non-relativistic} matter, to gravitational Maxwell-like equations
\cite{Maxwell-like equations}, which in turn lead to boundary conditions for
gravitational fields at the surfaces of the superconducting films homologous
to those of electromagnetism. All radiation fields, whether electromagnetic or
gravitational, will be treated classically, whereas the superconductors with
which they interact will be treated quantum mechanically. Thus, in this paper
we adopt a \emph{semi-classical} approach to the interaction of gravitational
radiation with matter.

Not enough effort has been made to investigate the ramifications of the
gravitational Maxwell-like equations for the interaction of GR waves with
matter, perhaps because the so-called \textquotedblleft electromagnetic
analogy\textquotedblright\ has been so hotly contested over the years
\cite{Kennefick}. In any case, we believe that these equations provide a helpful
framework for thinking about the response of non-relativistic matter to weak,
time-varying gravitational fields, especially that of macroscopically coherent
quantum charge and mass carriers, namely, the Cooper pairs of conventional,
type I superconductors. We argue here that the electromagnetic analogy
manifested in the Maxwell-like equations implies that type I superconductors
can be surprisingly efficient mirrors for GR waves at microwave frequencies.

In Section 2, we introduce the two basic claims upon which the larger argument
rests. Together, these two claims open the door to an enormously enhanced
interaction between a GR microwave and a type I superconductor, relative to
what one would expect in the case of a neutral superfluid or, indeed, any
normal metal or other classical matter. The first claim is that a GR microwave
will generate quantum probability supercurrents, and thus mass and electrical
supercurrents, inside a type I superconductor, due to the quantum mechanical
\emph{non-localizability }of the Cooper pairs within the material.

The non-localizability of Cooper pairs, which is ultimately due to the
Uncertainty Principle (UP), causes them to undergo \emph{non-picturable}%
,\emph{ non-geodesic} motion in the presence of a GR wave. This non-geodesic
motion, which is accelerated motion \emph{through }space, leads to the
existence of mass and charge supercurrents inside a superconductor. By
contrast, the localizability of the ions within the superconductor's lattice
causes them to undergo \emph{picturable}, \emph{geodesic }motion, i.e., free
fall, in the presence of the same wave. The resulting relative motion between
the Cooper pairs and the ionic lattice causes the electrical polarization of
the superconductor in the presence of a GR wave, since its Cooper pairs and
ions carry not only mass but oppositely signed charge as well.

Furthermore, the non-localizability of the Cooper pairs is \textquotedblleft
protected\textquotedblright\ from the normal process of localization, i.e.,
from decoherence, by the characteristic energy gap of the
Bardeen-Cooper-Schrieffer (BCS) theory of superconductivity. The decoherence
of entangled quantum systems such as Cooper pairs (which are in the spin-singlet state) is the fundamental cause of the \emph{localizability} of all
normal matter \cite{Zurek}. Indeed, this \textquotedblleft
classicalizing\textquotedblright\ process must occur within any spatially
extended system before the idea of the \textquotedblleft universality of free
fall\textquotedblright\ \cite{Adelberger} can be meaningfully applied to its
parts. After all, the classical principle behind the universality of free
fall, the Equivalence Principle (EP), is a strictly \emph{local} principle
\cite{WEP}.

The second of the two claims presented in Section 2 is that the mass
supercurrents induced by a GR wave are much stronger than what one would
expect in the case of a neutral superfluid or any normal matter, due to the
electrical polarization of the superconductor caused by the wave. This is what
we refer to as the \textquotedblleft Heisenberg-Coulomb\ (H-C)
effect.\textquotedblright\ The magnitude of the enhancement due to the H-C
effect (derived in Section 7) is given by the ratio of the electrical force to
the gravitational force between two electrons,%
\begin{equation}
\frac{e^{2}}{4\pi\varepsilon_{0}Gm_{e}^{2}}=4.2\times10^{42}\text{
,}\label{Ratio of electrical to gravitational forces}%
\end{equation}
where $e$ is the electron charge, $m_{e}$ is the electron mass, $\varepsilon
_{0}$ is the permittivity of free space, and $G$ is Newton's constant. The
enormity of (\ref{Ratio of electrical to gravitational forces}) implies the
possibility of an enormous back-action of a superconductor upon an incident GR
wave, leading to its reflection.

Of the four fundamental forces of nature, viz., the gravitational, the
electromagnetic, the weak, and the strong forces, only gravity and electricity
have long range, inverse square laws. The pure number obtained in
(\ref{Ratio of electrical to gravitational forces}) by taking the ratio of
these two inverse-square laws is therefore just as fundamental as the fine
structure constant. Because this number is so large, the gravitational force
is typically ignored in treatments of the relevant quantum physics. But as we
shall see below, a semi-classical treatment of the interaction of a
superconductor with a GR wave must account for both the electrodynamics and
the gravito-electrodynamics of the superconductor, since both play an
important role in its overall response to a GR wave.

In Section 3, we consider the interaction between an EM wave and a thin
metallic film having an arbitrary, frequency-dependent complex conductivity.
We determine the relevant boundary conditions using Faraday's and Ampere's
laws in order to derive general expressions for the transmissivity and
reflectivity of a thin film. In Section 4, we show that, in the case of a
superconducting film, the BCS theory implies that EM waves at microwave
frequencies will be specularly reflected even from films whose thickness is
less than the London penetration depth of the material, or, equivalently (at
sufficiently low frequencies), less than the material's plasma skin depth, as
has been experimentally observed \cite{Glover-and-Tinkham,Tinkham}. We
show, furthermore, that the frequency at which reflectivity drops to 50\%,
what we call the \textquotedblleft roll-off frequency\textquotedblright%
\ $\omega_{\text{r}}$, depends only on the ratio of the speed of light $c$ to
a single parameter, the length scale $l_{\text{k}}$ associated with the
kinetic inductance\ $L_{\text{k}}$ of the film's Cooper pairs
\cite{Merservey-and-Tedrow}, which in turn depends on the plasma skin depth
$\delta_{\text{p}}$. In the electromagnetic case, the microscopic size of
$\delta_{\text{p}}$ leads to a microscopic value for $l_{\text{k}}$ and thus
to the possibility of specular reflection over a wide range of frequencies
(including microwave frequencies) in the EM case.

In Section 5, we review the Maxwell-like equations for linearized Einsteinian
gravity and highlight the fact that any normal matter, with its inherently
high levels of dissipation, will necessarily be an inefficient reflector of GR
waves because of its high impedance relative to the extremely low
\textquotedblleft gravitational characteristic impedance of free
space\textquotedblright\ $Z_{\text{G}}$ ($2.8\times10^{-18}$ in SI
units).\ Superconductors, on the other hand, are effectively
\emph{dissipationless }at temperatures near absolute zero because of their
quantum mechanical nature \cite{Tinkham}. The fact that a superconductor's
effectively \emph{zero }impedance can be much smaller than the very small
quantity $Z_{\text{G}}$ allows it to reflect an incoming GR wave, much as a
low-impedance connection or \textquotedblleft short\textquotedblright\ at the
end of a transmission line can reflect an incoming EM wave.

In Section 6, we appeal to the Maxwell-like equations introduced in Section 5,
to the identicality of the boundary conditions that follow from them, and to
the linearity of weak GR-wave optics, in order to introduce GR analogs of the
earlier EM expressions for the reflectivity and roll-off frequency. As in the
EM case, the GR roll-off frequency $\omega_{\text{r,G}}$ can be expressed as
the ratio of the speed of light $c$ to a single parameter. In this case,
however, the relevant parameter is the length scale $l_{\text{k,G}}$
associated with the \emph{gravitational }kinetic inductance\ $L_{\text{k,G}}$
of the Cooper pairs. In this section we treat the superconductor as if it were
a neutral superfluid, i.e., as if its Cooper pairs were electrically neutral
particles interacting with one another and the with ionic lattice exclusively
through their mass. Although this assumption is unphysical, it leads to a
result in agreement with conventional wisdom, namely, that the gravitational
plasma skin depth $\delta_{\text{p,G}}$ and the kinetic inductance length
scale $l_{\text{k,G}}$ will be astronomical in size ($\sim$ $10^{13}$ m and
$\sim10^{36}$ m, respectively). Such enormous values imply that $\omega
_{\text{r,G}}$ will be effectively zero, and thus that superconductors cannot
function as mirrors for GR microwaves in laboratory-scale experiments.

In Section 7, we show why the approach taken at the end of the previous
section, in accord with conventional wisdom, is wrong. Superconductors
\emph{can }function as laboratory-scale mirrors for GR microwaves because of
the H-C effect. When one takes into account the electrical charge separation
induced within a superconductor by a GR wave (due to the BCS-gap-protected
non-localizability of its Cooper pairs), the ratio given in
(\ref{Ratio of electrical to gravitational forces}) enters into the analysis
in such a way as to keep $l_{\text{k,G}}$ microscopic and to raise
$\omega_{\text{r,G}}$ to the level of $\omega_{\text{r}}$. Thus the H-C effect
greatly enhances the reflection of a GR wave from the surface of a
superconductor -- by $42$ orders of magnitude! -- relative to what one would
expect from a neutral superfluid, a normal metal, or any normal matter.

Because both charge supercurrents and mass supercurrents are generated by an
incoming GR wave (and by an incoming EM wave), it is also necessary to
consider whether superconducting films are not mirrors but rather transducers,
i.e., converters, of GR radiation into EM radiation (in the case of an
incident GR wave), or vice versa (in the case of an incident EM wave). In
Section 8, we take up this particular question and show that transduction in
both directions is too weak to decrease reflection by any appreciable amount.
In section 9, however, we show that energy is conserved only when transduction
is included in the overall analysis as an effective absorption mechanism.

Finally, in Section 10 we indicate several possible experimental tests of the
basic claims advanced in the paper and offer brief comments on the broader
theoretical implications of our central hypothesis. Whereas present GR-wave
experiments aim to passively detect GR waves originating from astrophysical
sources, our argument implies the possibility of several new types of
laboratory-scale experiment involving GR waves. One type would test the physics
behind the Heisenberg-Coulomb effect by looking for a departure from geodesic
motion in the case of two coherently connected superconducting bodies that are
allowed to fall freely through a distance large enough to observe tidal
effects. A second type would investigate the existence and strength of any
gravitational Casimir-like effect between two type I superconductors. Yet a
third type, involving an electrically charged pair of superconductors, would
allow for more direct investigation of the existence and properties of
GR-waves, the results of which would bear significantly on the search for a
quantized theory of\ gravity.

Three appendices address ancillary issues: (A) the relationship between the
magnetic and kinetic inductances of a thin film, (B) the kinetic inductance
length scale according to a collisionless plasma model, and (C) the
relationship between the impedance argument given in Section 5 and Weinberg's
argument regarding the scattering cross-section of a Weber-style resonant bar
antenna, including an application of the Kramers-Kronig relations to the sum
rule for the strength of the interaction between a GR wave and a superconductor.

\section{The Uncertainty Principle limits the applicability of the Equivalence
Principle}

It is helpful to begin the analysis with a simple model of the interaction
between a weak GR wave and a normal metallic film. For the sake of eventually
considering the possibility of mirrors (i.e., the possibility of
\textquotedblleft ray\textquotedblright\ optics), we will assume here and
throughout that the lateral dimensions of the film are very large when
compared to the wavelength of the incident wave. Focusing on waves with very
high frequencies, i.e., microwaves, will allow us to treat the ions and normal
electrons of a laboratory-scale film as though they were freely floating,
non-interacting \textquotedblleft dust\textquotedblright\ or point particles
undergoing free fall along classical trajectories, i.e., traveling along geodesics.

Although it would be possible in principle, in this approximation, to detect
the passage of a GR wave over the film by observing the geodesic deviation
among its different components (the principle underlying LIGO), the film
\emph{cannot}, in this approximation,\emph{ }interact energetically with a
very high frequency GR wave. It cannot absorb or scatter any of the wave's
energy because each of its localized particles must, according to the EP,
travel along a geodesic, i.e., each particle must remain at rest with respect
to its local, co-moving, and freely-falling inertial frame \cite{Teukolsky}.
And since there can be no energetic interaction with the wave, mass currents
cannot be generated locally within the film without violating the conservation
of energy.

\begin{figure}[tbh]
\begin{center}
\includegraphics[
height=2.5in,
width=2.5in
]%
{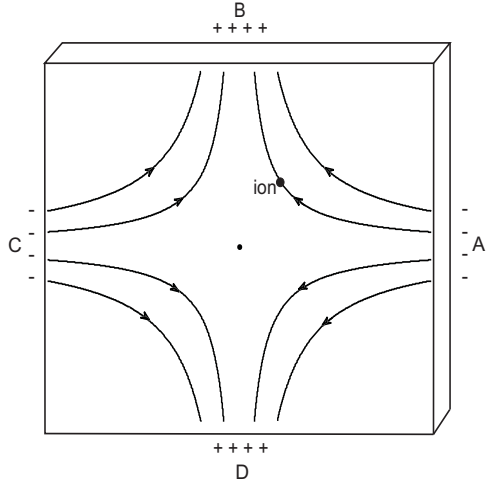}%
\caption{A
snapshot of a square metallic plate with a very high frequency GR wave
incident upon it at the moment when the gravitational tidal \textquotedblleft
forces\textquotedblright\ on the plate are those indicated by the hyperbolae,
as seen by a distant observer. All ions, being approximately in free fall, are
carried along \emph{with} space rather than accelerated \emph{through} space.
No work is done on them, and thus no kinetic energy is transferred to them, by
the wave. When the metal in the plate is normal, all ions and all normal
electrons locally co-move together along the same geodesics in approximate
free fall, so that the plate remains neutral and electrically unpolarized.
However, when the plate becomes superconducting, the Cooper pairs, being in
non-local \emph{entangled} states, remain at rest with respect to the center
of mass according to the distant observer, and do not undergo free fall along
with the ions and any residual normal electrons. This non-picturable,
non-geodesic, \emph{accelerated} motion of the Cooper pairs \emph{through
}space leads to \emph{picturable} quantum probability supercurrents, which
follow the same hyperbolae as the incident tidal GR wave fields (see Eqs.
(\ref{j with A and h})-(\ref{velocity field})). Since the Cooper pairs carry
not only mass but also charge, both mass and electrical supercurrents are
generated, and both types of current carry energy extracted from the
gravitational wave. In the snapshot shown, this leads to the accumulation of
positive charge at B and D, and to the accumulation of negative charge at
A and C, i.e., to a quadrupolar-patterned electrical polarization of the
superconductor. The resulting enormous Coulomb forces strongly oppose the
effect of the incoming tidal gravitational fields, resulting in the
mirror-like reflection of the incoming GR wave.}%
\label{fig:ion-trajectories}
\end{center}%
\end{figure}

It is true that a distant inertial observer will see the \textquotedblleft dust\textquotedblright\ particles undergo
quadrupolar motion, and will thus expect the film to emit GR radiation. But
this apparent paradox can be resolved by noting that the wave causes the
film's ions and normal electrons (which are to be treated as \emph{test}
particles whose masses and gravitational fields are negligible)\ to be carried
along\emph{ with }space\emph{ }rather than accelerated\emph{ through }space.
Only the latter kind of motion, in which the wave does work on the particles,
and hence transfers kinetic energy to them, leads to the time-varying mass
quadrupole moment that enters into Einstein's quadrupole formula for the
emission of GR radiation (see Figure \ref{fig:ion-trajectories}).

The classical concept of a \textquotedblleft geodesic\textquotedblright%
\ depends fundamentally upon the localizability,\ or spatial separability,\ of
particles. From a quantum mechanical point of view, localizability arises
ultimately from the \emph{decoherence of entangled states}, i.e., from the
\textquotedblleft collapse\textquotedblright\ of nonfactorizable
superpositions of product wavefunctions of two or more particles located at
two or more spatially well-separated points in space, into spatially
separable, factorizable, product wavefunctions, upon the interaction of the
particles with their environment. Decoherence typically occurs on extremely
short time-scales due to the slightest interaction with the environment
\cite{Zurek}. Whenever it does occur, one can speak classically of point
particles having trajectories or traveling along geodesics. Only \emph{after
}decoherence has occurred does the Equivalence Principle become a well-defined
principle, for only then does a particle's geodesic become well defined. In
other words, only through decoherence does the law of the \textquotedblleft
universality of free fall,\textquotedblright\ i.e., the experimentally
well-established claim that \textquotedblleft the gravitational acceleration
of a point body is independent of its composition\textquotedblright%
\ \cite{Adelberger}, become meaningful.

Entangled quantum states imply the nonlocalizability of particles, in the
sense that such states lead to experimentally well-confirmed violations of
Bell's inequalities \cite[Chapters 6 and 19]{Garrison-Chiao}. We claim here
that Cooper pairs are completely non-localizable within a superconductor, not
only in the sense of Heisenberg's Uncertainty Principle, but also because each
electron in a given Cooper pair in the BCS ground state is in an entangled
state, since each pair is in a superposition state of the product of two
electron wavefunctions with opposite momenta, and\emph{ }also simultaneously
in a superposition state of the product of two opposite electron spin-1/2
states (i.e., a spin-singlet state). The violation of Bell's inequalities by
these entangled states in the BCS ground state means that this state is
\textit{non-local}, in the sense that instantaneous correlations-at-a-distance
between the two electrons of a given Cooper pair must occur in the
superconductor upon remote measurements within a long, single continuous piece
of superconductor (the distance between these remote measurements can be
arbitrarily large). Although these instantaneous correlations-at-a-distance
cannot be used to send signals faster than light \cite{Garrison-Chiao}, they
also cannot be accounted for in any local, realistic theory of quantum
phenomena, including those which satisfy the completeness conditions demanded
by Einstein, Podolsky, and Rosen (EPR) \cite{EPR}.

The localizability or spatial separability of \emph{all }particles, as
envisioned by EPR, would of necessity lead to the \emph{universal validity} of
the Equivalence Principle, and thus to the idea that even Cooper pairs must
undergo geodesic motion (i.e., free-fall) within a superconductor in response
to an incident GR wave. There could be no relative motion between the Cooper
pairs and the ions, no spatial separation of charges inside the
superconductor, and no enhancement, even in principle, of the superconductor's
interaction with a GR wave relative to that of a normal metal interacting
with the same wave. But Cooper pairs are manifestly \emph{not }localizable
within the superconductor, since they are fully quantum mechanical, non-local
systems. For this reason the \textquotedblleft
dust-particles-following-geodesics\textquotedblright\ model introduced earlier must fail in the case of a superconductor, even as a first approximation
\cite{wrong-interp-of-EP}.

When a conventional, type I superconductor is in the BCS ground state, each of
its Cooper pairs is in a zero-momentum eigenstate relative to the center of
mass of the system. According to Heisenberg's Uncertainty Principle (UP), the
fact that the Cooper pairs' momenta are perfectly definite entails that their
positions within the superconductor are completely \emph{uncertain}, i.e.,
that the pairs are \emph{non-localizable}. The motion of a given Cooper pair
within the superconductor is irreducibly quantum mechanical in nature, being
related to the pair's wavefunction. Such motion \emph{cannot }be pictured in
terms of a well-defined trajectory or geodesic \cite{Davies}. Indeed, at a
conceptual level, the ascription of a \textquotedblleft
trajectory\textquotedblright\ or \textquotedblleft geodesic\textquotedblright%
\ to a given Cooper pair within a superconductor becomes \emph{meaningless} in
the BCS ground state. This is similar to what Bohr taught us concerning the
meaninglessness of the concept of \textquotedblleft orbit\textquotedblright%
\ in the ground state of the hydrogen atom during its interaction with
radiation fields \cite[pp. 113ff]{Bohr}.

The robustness of the BCS ground state in the face of perturbations is
guaranteed by the BCS energy gap, which \textquotedblleft
protects\textquotedblright\ the Cooper pairs from making quantum transitions
into excited states, such as happens in pair-breaking (as long as the material
is kept well below its transition temperature and the frequency of the
incident radiation is below the BCS gap frequency \cite{pair-breaking}). The
energy gap prevents the pairs from decohering, and from becoming localized
like the superconductor's ions and any residual, normal conduction electrons
\cite{delocalization}. If the Cooper pairs cannot be thought of as
\emph{localizable} point bodies, then the \textquotedblleft
universality\textquotedblright\ of free fall cannot be meaningfully applied to
them. In short, an application of the EP to the motion of Cooper pairs within
a superconductor is fundamentally \emph{precluded} by the UP. This is not to
make the well-known point that quantum field theories may lead to measurable
\textquotedblleft quantum violations of the EP\textquotedblright\ due to
possible \textquotedblleft fifth-force\textquotedblright\ effects that produce
slight corrections to particle geodesics (see, for example, Adelberger
\cite{Adelberger} and Ahluwalia \cite{Ahluwalia}), but rather to observe that
the non-localizability of quantum objects places a fundamental limit on the
\emph{applicability} of the EP (a point previously raised by Chiao \cite[esp.
Section V]{Chiao-Wheeler}).

In contrast to a superconductor's non-localizable Cooper pairs, its ions
(and, at finite temperatures, any residual background of normal electrons) are
unaffected by the energy gap, and are thus fully \emph{localized }by the
decohering effect of their interactions with the environment. Thus, unlike Cooper pairs, the ionic lattice possesses no coherent quantum phase anywhere. The \emph{geodesic} motion of the ions will therefore differ from the
\emph{non-geodesic} motion of the Cooper pairs. The latter, which is accelerated motion \emph{through }space, implies the existence of quantum probability supercurrents, and thus of mass and electrical supercurrents, inside the superconductor (see Figure \ref{fig:ion-trajectories}). These supercurrents will carry energy extracted from the GR wave. The possibility of a non-negligible energetic interaction between a GR wave and a superconductor depends crucially upon this initial claim, which is implied by the absence of the localizing effect of decoherence upon the Cooper pairs.

Before we turn to the second claim, it is worth noting that the non-geodesic
motion of a superconductor's Cooper pairs also follows from what London called
the \textquotedblleft rigidity of the wavefunction\textquotedblright%
\ \cite{London}. The phase of the wavefunction of each Cooper pair must be
constant in the BCS ground state prior to the arrival of a GR wave. This
implies that the gradient of its phase is initially zero. Since an incoming GR
wave whose frequency is less than the BCS gap frequency cannot alter this
phase (in the lowest order of time-dependent perturbation theory), and since the canonical
momentum of any given pair relative to the center of mass of the
superconductor is proportional to the gradient of its phase, the canonical
momentum of each pair must remain zero at all times with respect to the center
of mass of the system in the presence of a GR wave, as seen by the distant
inertial observer.

This quantum-type rigidity implies that Cooper pairs will acquire kinetic
energy from a GR wave in the form of a nonzero \emph{kinetic} velocity, i.e., that
they will be \emph{accelerated} by the wave relative to any local inertial
frame whose origin does not coincide with the center of mass of the system
(for example, at the corners of a large, square superconducting film; see
Section 7). In other words, the apparent \textquotedblleft
motionlessness\textquotedblright\ of the Cooper pairs in the presence of a GR
wave, as witnessed by a distant inertial observer, in fact entails their
accelerated motion \emph{through} local space. Again, this behavior implies
the existence of mass supercurrents inside the superconductor that carry
energy extracted from the wave.

Of course, even normal matter such as in a Weber-style resonant bar detector
has some extremely small degree of rigidity arising from its very weak
interatomic coupling. Thus normal matter does not, strictly speaking, behave
as a collection of freely falling, noninteracting \textquotedblleft dust
particles\textquotedblright\ in the presence of a very low frequency GR
wave.\ Instead, like the Cooper pairs, but to a much smaller degree, and at
much lower frequencies than the microwave frequencies being considered here,
normal matter opposes the squeezing and stretching of space going on around it
(as Feynman pointed out in his well-known remarks on why GR waves must carry
energy \cite{Feynman}). Thus, even normal matter will acquire an extremely
small amount of kinetic energy as it is accelerated through space by a passing
GR wave. In this case, though, high levels of dissipation inside the material
will cause whatever small amount of energy is extracted from the GR wave to be
overwhelmingly converted into heat instead of being predominantly re-radiated
as a scattered GR wave (as Weinberg has pointed out \cite{Weinberg}). A key
feature of the mass supercurrents carried by Cooper pairs is that they are
\emph{dissipationless}. We shall return to this particular point in Section 5.

The second basic claim underlying the paper's larger argument follows from the
dual nature of the supercurrents generated by a GR wave within a
superconductor. Since a GR wave will generate both mass and charge
supercurrents, it will \emph{electrically polarize }the superconductor. This
important observation implicates the Coulomb force of attraction between the
oppositely signed charges that must accumulate at the edges of the
superconductor, if there is to be no violation of charge conservation (see
Figure \ref{fig:ion-trajectories}). These oppositely signed charges will
consist of negatively charged Cooper pairs, on the one hand, and
corresponding, positively charged Cooper-pair holes\ (hereafter,
\textquotedblleft holes\textquotedblright), on the other. An incoming GR wave
with a frequency well below the superconductor's plasma frequency will thus
generate a virtual plasma excitation inside the superconductor. The resulting
\emph{Coulomb }force between the Cooper pairs and holes, which acts as a
Hooke's law restoring force, strongly opposes the effect of the incident wave.
The enormous back-action of this force on the motion of the Cooper pairs
greatly enhances their \emph{mass }conductivity (see Section 7), to the point
where specular reflection of an incident GR wave from a superconducting film
becomes possible. The existence of strengthened mass supercurrents within a
superconductor, which is due to the combined effect of the quantum
non-localizability of the Cooper pairs and the Coulomb attraction between the
pairs and holes, is what we refer to as the \textquotedblleft
Heisenberg-Coulomb effect.\textquotedblright

Consider, by way of contrast, what happens when a GR wave impinges on a
superfluid, whose constituent particles are electrically \emph{neutral}. Mass
supercurrents will again be induced by the wave, due to quantum
non-localizability, but in this case there will be no enhancement effect
because the mass carriers within a superfluid are its electrically neutral atoms.
Thus no appreciable fraction of incident GR-wave power can be reflected from
the surface of a neutral superfluid. On the other hand, one might worry that
the size of the H-C effect in a superconductor would drive its mass
supercurrents above the critical level, thereby undermining the possibility of
specular reflection. But it should always be possible to arbitrarily reduce
the amplitude of the driving radiation field until the superconductor responds
\emph{linearly }to the field (see the related discussion of superluminality at
the end of Section 7). The existence of a linear-response regime guarantees
the possibility of fabricating \emph{linear} GR-wave optical elements,
including mirrors.

\section{The interaction of an EM wave with a thin metallic film}

The question of the interaction of an EM wave with a metallic film whose
thickness $d$ is small compared to the wavelength can be addressed using
\textquotedblleft lumped-circuit\textquotedblright\ concepts such as
resistance, reactance, inductance, etc., of an infinitesimal square element of
the film. (As before, we assume, for the sake of considering mirror-like
behavior, that the lateral dimensions of the film are at least on par with the
wavelength of the incident wave.) In this section we derive a formula for the
transmissivity ${\mathcal{T}}$ as well as the reflectivity ${\mathcal{R}}$ of
a thin metallic film with an arbitrary, frequency-dependent complex
conductivity. In the next section we apply this analysis to the case of a
\emph{superconducting }film.

\begin{figure}
[ptb]
\begin{center}
\includegraphics[
height=2.2in,
width=3.2in
]%
{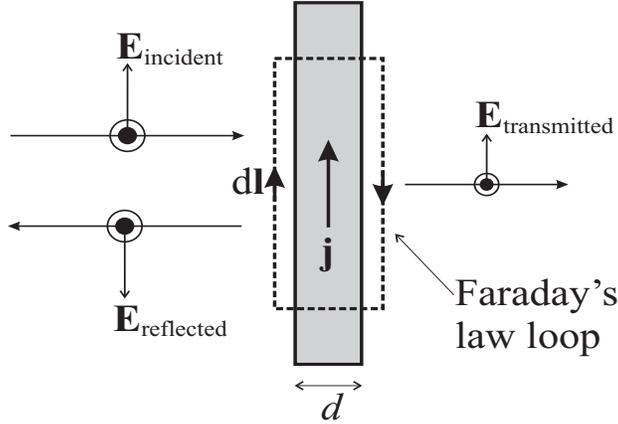}%
\caption{A thin
metallic film of thickness $d$ is straddled by a rectangular loop (dashed
lines) for applying Faraday's law to it. An incident EM wave is partially
transmitted and partially reflected by the film. The EM wave generates an
electrical current density \textbf{j}, which flows uniformly inside the film.
A similar rectangular loop (not shown) lying in a plane parallel to the
magnetic fields (denoted by the circles with central dots) is for applying
Ampere's law.}%
\label{fig:thin-metallic-film}%
\end{center}
\end{figure}

The complex amplitude reflection coefficient $r$ corresponding to the
proportion of incident EM radiation at frequency $\omega$ reflected from a
thin film and the complex amplitude transmission coefficient $t$ corresponding
to the proportion of the same radiation transmitted through the film can be
defined as follows:%
\begin{subequations}
\begin{align}
\mathbf{E}_{\text{reflected}}  &  =r\mathbf{E}_{\text{incident}}\\
\mathbf{E}_{\text{transmitted}}  &  =t\mathbf{E}_{\text{incident}}\text{ }.
\end{align}
By convention, $r,$ if it is real, is defined to be $positive$ when the
reflected electric field $\mathbf{E}_{\text{reflected}}$ is $oppositely$
directed to the incident electric field $\mathbf{E}_{\text{incident}}$. On the
other hand, $t,$ if it is real, is defined to be $positive$ when the
transmitted electric field $\mathbf{E}_{\text{transmitted}}$ points in the
$same$ direction as the incident electric field $\mathbf{E}_{\text{incident}}%
$. In general, $r$ and $t$ are complex quantities whose values depend on the
frequency $\omega$ of the incident wave, but all radiation fields will be
treated classically.

Since the tangential components of the electric fields must be continuous
across the vacuum-film interface, the electric field inside the film
$\mathbf{E}_{\text{inside}}$\ drives a current density $\mathbf{j}$ inside the
film that is \emph{linearly }related to this driving electric field, for the
general case of a \emph{linear-response }theory of the interaction of matter
with \emph{weak }driving fields. This linear relationship is given by%
\end{subequations}
\begin{equation}
\mathbf{j(}\omega\mathbf{)=\sigma}(\omega)\mathbf{E}_{\text{inside}}%
(\omega)\text{, where} \label{j=sigma-E}%
\end{equation}%
\begin{equation}
\mathbf{E}_{\text{inside}}=(1-r)\mathbf{E}_{\text{incident}}\text{ \ at
frequency }\omega. \label{E-inside}%
\end{equation}
In general, the conductivity $\mathbf{\sigma}(\omega)$ associated with the
current generated within the film at a given driving frequency $\omega$ will
be a complex quantity:%
\begin{equation}
\sigma(\omega)=\sigma_{1}(\omega)+i\sigma_{2}(\omega),
\end{equation}
where $\sigma_{1}(\omega)$ represents the current's in-phase, dissipative
response at frequency $\omega$ to the driving field at frequency $\omega$, and
$\sigma_{2}(\omega)$ represents the current's out-of-phase, non-dissipative
response at the same frequency \cite{conductivity-sign-convention}.

If the thickness of the film $d$ is much less than a wavelength of the
incident radiation, then the right-hand side of Faraday's law applied to the
loop shown in Figure \ref{fig:thin-metallic-film} encloses a negligible amount
of magnetic flux $\Phi_{B}$, so that%
\begin{equation}%
{\displaystyle\oint}
\mathbf{E}\cdot d\mathbf{l}=\mathbf{-}\frac{d\Phi_{B}}{dt}\rightarrow0\ .
\end{equation}
Using the sign conventions introduced above, one finds that%
\begin{equation}
1-r-t=0\text{ .} \label{1st-r-t-relation}%
\end{equation}

Now let us apply Ampere's law \cite{Sign-of-mass-current}%
\begin{equation}%
{\displaystyle\oint}
\mathbf{H}\cdot d\mathbf{l}=I \label{Ampere's-law}%
\end{equation}
to the \textquotedblleft Amperian\textquotedblright\ loop (not shown in Figure
\ref{fig:thin-metallic-film}) whose plane is parallel to the magnetic fields
of the incident, reflected, and transmitted EM waves, and perpendicular to the
Faraday's law loop shown in Figure \ref{fig:thin-metallic-film}. Let this
Amperian loop span the entire width $w$ of the film in the direction of the
magnetic field. For a plane EM wave propagating in free space,%
\begin{equation}
\left\vert \mathbf{H}\right\vert =\frac{\left\vert \mathbf{B}\right\vert }%
{\mu_{0}}=\frac{\left\vert \mathbf{E}\right\vert }{Z_{0}}\text{ ,}%
\end{equation}
where $Z_{0}$ is the characteristic impedance of free space and $\mu_{0}$ is
the magnetic permeability of free space. It then follows that%
\begin{equation}
w(1+r-t)\frac{E_{\text{incident}}}{Z_{0}}=I\text{ ,}%
\end{equation}
where%
\begin{equation}
I=Aj=\sigma wd(1-r)E_{\text{incident}}%
\end{equation}
is the total enclosed current being driven inside the film by the applied
electric field inside the film (\ref{E-inside}), which leads to%
\begin{equation}
\frac{1}{Z_{0}}\left(  1+r-t\right)  =\sigma d(1-r)\text{ .}
\label{2nd-r-t-relation}%
\end{equation}

From (\ref{1st-r-t-relation}) and (\ref{2nd-r-t-relation}) we have two
equations in the two unknowns $r$ and $t$, which can be rewritten as%
\begin{subequations}
\begin{align}
1-r-t  &  =0\text{ , and}\\
1+r-t  &  =x(1-r)\text{ ,}%
\end{align}
where $x\equiv\sigma Z_{0}d$ . Solving for $1/t$ and $\,1/r$, one obtains%
\end{subequations}
\begin{subequations}
\begin{align}
\frac{1}{t}  &  =1+\frac{1}{2}x,\text{ and}\label{1/t}\\
\frac{1}{r}  &  =1+2\frac{1}{x}. \label{1/r}%
\end{align}

Using the definition ${\mathcal{T}}=tt^{\ast}=\left\vert t\right\vert ^{2},$
one then obtains for the reciprocal of the transmissivity%
\end{subequations}
\begin{align}
\frac{1}{{\mathcal{T}}}  &  =\frac{1}{tt^{\ast}}=\left(  1+\frac{1}%
{2}x\right)  \left(  1+\frac{1}{2}x^{\ast}\right) \\
&  =1+\frac{1}{2}\left(  x+x^{\ast}\right)  +\frac{1}{4}xx^{\ast}\nonumber\\
&  =1+\operatorname{Re}x+\frac{1}{4}\left\{  (\operatorname{Re}x)^{2}+\left(
\operatorname{Im}x\right)  ^{2}\right\} \nonumber\\
&  =\left(  1+\frac{1}{2}\operatorname{Re}x\right)  ^{2}+\frac{1}{4}\left(
\operatorname{Im}x\right)  ^{2}\text{ .}\nonumber
\end{align}
Substituting $x=\sigma Z_{0}d=(\sigma_{1}+i\sigma_{2})Z_{0}d$ into this
expression, one finds that%
\begin{equation}
{\mathcal{T}}=\left\{  \left(  1+\frac{1}{2}\sigma_{1}Z_{0}d\right)
^{2}+\left(  \frac{1}{2}\sigma_{2}Z_{0}d\right)  ^{2}\right\}  ^{-1}\text{ .}%
\end{equation}
This general result, which applies to \emph{any} thin metallic film with a
complex conductivity, agrees with Tinkham's expression for ${\mathcal{T}}$
\cite[Eq. (3.128)]{Tinkham} in the case of a superconducting film when the
index of refraction of the film's substrate in his expression is set equal to
unity (i.e., when the film is surrounded on \emph{both }sides by free space).

Similarly, using the definition ${\mathcal{R}}=rr^{\ast}=\left\vert
r\right\vert ^{2},$ one obtains for the reciprocal of the reflectivity%
\begin{align}
\frac{1}{{\mathcal{R}}}  &  =\frac{1}{rr^{\ast}}=(1+2y)(1+2y^{\ast})\\
&  =1+2(y+y^{\ast})+4yy^{\ast}\nonumber\\
&  =1+4\operatorname{Re}y+4\left\{  (\operatorname{Re}y)^{2}+\left(
\operatorname{Im}y\right)  ^{2}\right\} \nonumber\\
&  =\left(  1+2\operatorname{Re}y\right)  ^{2}+4\left(  \operatorname{Im}%
y\right)  ^{2}\text{ ,}\nonumber
\end{align}
where%
\begin{equation}
y\equiv\frac{1}{x}=\frac{1}{\sigma Z_{0}d}=\frac{\rho}{Z_{0}d}%
\end{equation}
and $\rho$ is the complex resistivity of the film (again at frequency $\omega
$). In general, $\rho$ and $\sigma$ are related by%
\begin{equation}
\rho\equiv\frac{1}{\sigma}=\frac{1}{\sigma_{1}+i\sigma_{2}}=\frac{\sigma
_{1}-i\sigma_{2}}{\sigma_{1}^{2}+\sigma_{2}^{2}}=\rho_{1}+i\rho_{2}\text{ ,}%
\end{equation}
where%
\begin{subequations}
\begin{align}
\rho_{1}  &  =\frac{\sigma_{1}}{\sigma_{1}^{2}+\sigma_{2}^{2}}\\
\rho_{2}  &  =-\frac{\sigma_{2}}{\sigma_{1}^{2}+\sigma_{2}^{2}}\text{ .}%
\end{align}
The reflectivity of any thin metallic film with complex conductivity is
therefore%
\end{subequations}
\begin{align}
{\mathcal{R}}  &  =\left\{  \left(  1+2\frac{\sigma_{1}}{\sigma_{1}^{2}%
+\sigma_{2}^{2}}\frac{1}{Z_{0}d}\right)  ^{2}\right. \nonumber\\
&  \quad\;\left.  +\left(  2\frac{\sigma_{2}}{\sigma_{1}^{2}+\sigma_{2}^{2}%
}\frac{1}{Z_{0}d}\right)  ^{2}\right\}  ^{-1}\text{ .} \label{EM-reflectivity}%
\end{align}
Although the precise degree of reflection for a film of given thickness $d$
will depend on the specific character of the film's conductivity, the presence
of the sum inside the first squared term of (\ref{EM-reflectivity}) indicates
that the dissipative component of the conductivity $\sigma_{1}$ will inhibit
reflection more strongly than the non-dissipative component $\sigma_{2}$. With
this clear hint of the importance of dissipationlessness for achieving
specular reflection, we turn our attention to superconducting\emph{ }films.

\section{A criterion for the specular reflection of EM waves from
superconducting films}

The BCS theory of superconductivity has been confirmed by many experiments.
Here we review the application of this well-established theory to the problem
of mirror-like reflection of EM waves from a superconducting film. We consider
once again a film whose thickness $d$ is small enough to make the use of
\textquotedblleft lumped-circuit\textquotedblright\ concepts legitimate, but
which is now also much smaller than the coherence length $\xi_{0}$ and the
London penetration depth $\lambda_{\text{L}}$ of the material (i.e., the
so-called \textquotedblleft local\textquotedblright\ or \textquotedblleft
dirty\textquotedblright\ limit).

As Tinkham has noted \cite[p. 39]{Tinkham}, the dissipative part of the
conductivity of such a film $\sigma_{1\text{s}}$ goes \emph{exponentially} to
zero as $T\rightarrow0$ in response to a driving wave whose frequency is less
than%
\begin{equation}
\omega_{\text{gap}}=\frac{2\Delta(0)}{\hbar}\cong\frac{3.5k_{\text{B}%
}T_{\text{c}}}{\hbar}\text{ ,} \label{BCS critical temperature}%
\end{equation}
where $\Delta(0)$ (henceforward abbreviated as $\Delta)$ is the gap energy per
electron of the BCS theory at $T=0$, $k_{\text{B}}$ is Boltzmann's constant,
and $T_{\text{c}}\,$is critical temperature for the superconducting
transition. The exponential suppression of the film's dissipative response is
due to the \textquotedblleft freezing out\textquotedblright\ of its normal
electrons through the Boltzmann factor $\exp\left(  -\Delta/k_{\text{B}%
}T\right)  $ as $T\rightarrow0$.

On the other hand, the film's non-dissipative conductivity $\sigma_{2\text{s}%
}$ rises asymptotically to some finite value in the same limit \cite[Eq.
(3.125)]{Tinkham}. The behavior of $\sigma_{2\text{s}}$, which can be
calculated using the BCS theory, is due to the film's inductive reactance
$X_{\text{L}}$, which in turn arises from its inductance (per square element
of the film) $L$. These three parameters are related to one another by%
\begin{equation}
\frac{1}{\sigma_{2\text{s}}d}=X_{\text{L}}=\omega L\text{ .}
\label{inductive-reactance}%
\end{equation}
For a superconducting film at temperatures sufficiently near $T=0$ (e.g., in
the milli-Kelvin range for a Pb film) and for frequencies lower than
$\omega_{\text{gap}}$, the ohmic dissipation of the film will be exponentially
suppressed by the Boltzmann factor, so that one can, to a good approximation,
set $\sigma_{1\text{s}}=0$ and rewrite (\ref{EM-reflectivity}) as
\cite{transmission-line-reflectivity}:%
\begin{equation}
{\mathcal{R}}_{\text{s}}=\left\{  1+\left(  2\frac{X_{\text{L}}}{Z_{0}%
}\right)  ^{2}\right\}  ^{-1}\text{ .} \label{EM-reflectivity-imped-ratio}%
\end{equation}

The two previous expressions allow us to define an \textquotedblleft upper
roll-off frequency\textquotedblright\ $\omega_{\text{r}}$ for the reflection
of EM waves from a superconducting film, i.e., the frequency at which
reflectivity drops to $50\%$ (when the film is kept at nearly $T=0$ and when
$\omega<\omega_{\text{gap}}$):%
\begin{equation}
\omega_{\text{r}}=\pm\frac{Z_{0}}{2L}\text{ ,} \label{Impedance-condition}%
\end{equation}
where we discard the negative solution as being unphysical. The film's lower
roll-off frequency\ is simply determined by its lateral dimensions, which for
mirror-like behavior to occur must be, as noted before, much larger than the
wavelength $\lambda=\left(  2\pi c\right)  /\omega$ of the incident EM wave.
Because the upper roll-off frequency is our primary concern, we refer to it
throughout\emph{ }as \emph{the }roll-off frequency. Unlike the lower roll-off
frequency, it depends on the intrinsic properties of the material and cannot
be adjusted at will by altering the lateral dimensions of the film.

The physical meaning of the expression for $\omega_{\text{r}}$ given in
(\ref{Impedance-condition}) is that a superconducting film whose dissipative
conductivity has been exponentially frozen out can \textquotedblleft short
out\textquotedblright\ and thus specularly reflect an incoming EM wave whose
frequency is below $\omega_{\text{gap}}$ \emph{as long as the film's
inductance is sufficiently small to allow non-dissipative supercurrents to
flow at frequencies less than }$\omega_{\text{gap}}$. As happens with an RF
choke, a large inductance will prevent supercurrents from being established
inside the film. Thus, the roll-off frequency and reflectivity will be lowered
to levels on par with those of a normal metal.

From (\ref{Impedance-condition}) it is clear that the possibility of specular
reflection of EM waves by a superconducting\emph{ }film at low temperatures
and frequencies depends crucially on the film's inductance $L$. The inductance
will have two components: a magnetic inductance\ $L_{\text{m}}$ due to the
magnetic fields created by the charge supercurrents carried by the Cooper
pairs, and a kinetic inductance\ $L_{\text{k}}$ due to the inertial mass of
the same Cooper pairs, which causes them to oppose the accelerating force of
the external electric field \cite[pp. 88, 99]{Tinkham}%
\cite{Merservey-and-Tedrow}. As it happens, $L_{\text{m}}$ is numerically
negligible compared to $L_{\text{k}}$ for a thin film (see Appendix A), so
that we can proceed under the assumption that $L\cong L_{\text{k}}$.

When $T$ $\ll T_{\text{c}}$ and $\omega\ll\omega_{\text{gap}}$, the BCS theory
yields the following relation between the imaginary part of a superconducting
film's complex conductivity $\sigma_{\text{2s}}$ and its normal conductivity
$\sigma_{\text{n}}$ \cite[Eq. (3.125a)]{Tinkham}:%
\begin{equation}
\sigma_{\text{2s}}=\frac{\pi\Delta}{\hbar\omega}\sigma_{\text{n}}\text{ .}%
\end{equation}
From the Drude model of metallic conductivity, it follows\textbf{ }\cite{Drude
Model} that a film of thickness $d$ will have a normal conductivity
$\sigma_{\text{n}}$ given by%
\begin{equation}
\sigma_{\text{n}}=\frac{n_{\text{e}}e^{2}d}{m_{\text{e}}v_{\text{F}}}\text{ ,}
\label{normal-conductivity}%
\end{equation}
where $e$ is the charge of the electron, $m_{\text{e}}$ is its mass,
$v_{\text{F}}$ is its Fermi velocity, and $n_{\text{e}}$ is the number density
of conduction electrons. Then $\sigma_{\text{2s}}$ becomes%
\begin{equation}
\sigma_{\text{2s}}=\frac{\pi\Delta}{\hbar\omega}\cdot\frac{n_{\text{e}}e^{2}%
d}{m_{\text{e}}v_{\text{F}}}\text{ ,} \label{sigma_2s}%
\end{equation}
from which it follows that the kinetic inductance can be expressed as%
\begin{equation}
L_{\text{k}}=\frac{1}{\omega\sigma_{2\text{s}}d}=\frac{1}{d^{2}}\cdot
\frac{\hbar v_{\text{F}}}{\pi\Delta}\cdot\frac{m_{\text{e}}}{n_{\text{e}}%
e^{2}}\text{ .} \label{L_k-details}%
\end{equation}
The $1/d^{2}$ term in (\ref{L_k-details}) indicates a dependence on the film's
thickness, whereas the presence of $\hbar v_{\text{F}}/\pi\Delta$ implies an
additional dependence on the coherence length $\xi_{0},$ since according to
the BCS theory%
\begin{equation}
\xi_{0}=\frac{\hbar v_{\text{F}}}{\pi\Delta}\text{ .} \label{coh-len}%
\end{equation}
The $m_{\text{e}}/n_{\text{e}}e^{2}$ term could be interpreted as the London
penetration depth $\lambda_{\text{L}}$, since%
\begin{equation}
\mu_{0}\lambda_{\text{L}}^{2}=\frac{m_{\text{e}}}{n_{\text{e}}e^{2}}\text{ .}
\label{lon-pen-dep}%
\end{equation}
However, in the present context it is more appropriate to relate this term to
the plasma frequency $\omega_{\text{p}}$ by%
\begin{equation}
\mu_{0}\frac{c^{2}}{\omega_{\text{p}}^{2}}=\frac{m_{\text{e}}}{n_{\text{e}%
}e^{2}}\text{ ,} \label{plas-freq}%
\end{equation}
since the Cooper pairs within a superconductor can be regarded as a type of
quantum mechanical, collisionless plasma \cite{Buisson}. We are, after all,
concerned not with the screening of DC magnetic fields through the Meissner
effect, but with the reflection of EM radiation -- with an
\emph{electrodynamic }effect rather than a \emph{magnetostatic }one. In the
limit of $\omega\ll\omega_{\text{p}},$ the plasma skin depth $\delta
_{\text{p}}$ (the depth to which an EM wave with a frequency $\omega$ can
penetrate into a plasma) is simply%
\begin{equation}
\delta_{\text{p}}=\frac{c}{\omega_{\text{p}}}\text{ ,}%
\end{equation}
so that in this limit%
\begin{equation}
\mu_{0}\delta_{\text{p}}^{2}=\frac{m_{\text{e}}}{n_{\text{e}}e^{2}}\text{ }.
\label{plas-skin-dep}%
\end{equation}
Comparing (\ref{plas-skin-dep}) with (\ref{lon-pen-dep}), we see that the
electrodynamic concept of the plasma skin depth and the magnetostatic limit
given by the London penetration depth coincide not just in the stronger limit
of $\omega\rightarrow0$ but also in the weaker limit of $\omega\ll
\omega_{\text{p}}$.

In light of these considerations, we can re-express the kinetic inductance
$L_{\text{k}}$ (\ref{L_k-details}) in terms of the permeability of free space
$\mu_{0}$, the coherence length $\xi_{0}$, the plasma skin depth
$\delta_{\text{p}}$, and the thickness of the film $d$:%
\begin{equation}
L_{\text{k}}=\mu_{0}\xi_{0}\left(  \frac{\delta_{\text{p}}}{d}\right)
^{2}\text{ .} \label{L_k-multiple-lengths}%
\end{equation}
It is then possible to express $L_{\text{k}}$ in more familiar form, i.e., as
the product of the magnetic permeability of free space and the kinetic
inductance length scale $l_{\text{k}}$:%
\begin{equation}
L_{\text{k}}=\mu_{0}l_{\text{k}}\text{ ,} \label{L_k-single-length}%
\end{equation}
where $l_{\text{k}}$ is%
\begin{equation}
l_{\text{k}}=\xi_{0}\left(  \frac{\delta_{\text{p}}}{d}\right)  ^{2}\text{ .}
\label{l_k}%
\end{equation}
(For a comparison of this BCS-based\ derivation of $l_{\text{k}}$ with one
based on plasma concepts, see Appendix B.)

We can now rewrite the film's inductive reactance $X_{\text{L}}$ in terms of
the frequency of the incident EM wave $\omega$, the permeability of free space
$\mu_{0}$, and the kinetic inductance length scale $l_{\text{k}}$:%
\begin{equation}
X_{\text{L}}=\omega L_{\text{k}}=\omega\mu_{0}l_{\text{k}}\text{ .}
\label{BCS-ind-react}%
\end{equation}
Returning to the crucial ratio of the inductive reactance to the
characteristic impedance of free space given earlier in
(\ref{Impedance-condition}), we see that the roll-off frequency becomes%
\begin{equation}
\omega_{\text{r}}=\frac{Z_{0}}{2L_{\text{k}}}=\frac{\mu_{0}c}{2\mu
_{0}l_{\text{k}}}=\frac{c}{2l_{\text{k}}}\text{ .} \label{roll-off-freq-BCS}%
\end{equation}
Notice that $\mu_{0}$ cancels out of the numerator and denominator of this
expression, so that the specular reflection of an EM wave with frequency
$\omega$ from a superconducting film at temperatures sufficiently near $T=0$
depends only on the ratio of the speed of light $c$ to the kinetic inductance
length scale $l_{\text{k}}$.

To make this claim concrete, let us consider here (and in subsequent examples)
the case of a thin lead (Pb) film with a thickness of $d=2$ nm and an angular
frequency for the incident radiation of $\omega=2\pi\times(6$ GHz$)$. The
known values for the coherence length and the London penetration depth of Pb
are $\xi_{0}=83$ nm and $\delta_{\text{p}}=\lambda_{\text{L}}=37$ nm,
respectively \cite[p. 24]{DeGennes}. Inserting these values into (\ref{l_k}),
we see that $l_{\text{k}}\approx30$ $\mu$m and, from (\ref{roll-off-freq-BCS}%
), that $\omega_{\text{r}}\approx2\pi\times(800$ GHz$)$. When we recall that
the theoretically calculated gap frequency for superconducting Pb at $T=0$ is
approximately $2\pi\times(500$ GHz$)$, we see that our estimate of
$\omega_{\text{r}}$ is roughly equivalent to the claim that $\omega
<\omega_{\text{gap}}$ for specular reflection to occur, which is consistent
with previously stated assumptions (and with the requirement that $\omega
\ll\omega_{\text{p}}$, since $\omega_{\text{p}}\approx2\pi\times$ $(1.3$
PHz$)$ for Pb).

The analysis presented in this section is in basic agreement with the
experiments of Glover and Tinkham \cite{Glover-and-Tinkham}, and it belies the
commonly held misconception that specular reflection can occur only when the
thickness of the material $d$ is greater than its skin depth $\delta
_{\text{p}}$ (or penetration depth $\lambda_{\text{L}}$). Reflection from a
superconducting film is due not to the gradual diminishment of the radiation
field as it enters the film but to the destructive interference between the
incident radiation and the radiation emitted in the forward scattering
direction by the sheet supercurrents set up within the film. In fact, a closer
examination of (\ref{l_k}) and (\ref{roll-off-freq-BCS}) reveals that
appreciable reflection of a 6 GHz EM wave can occur from a Pb film -- a type I
superconductor -- even when the film's thickness is as much as $2$ orders of
magnitude smaller than its characteristic penetration depth. A type II
superconductor, on the other hand, will generate considerable losses, due to
the ohmic or dissipative flux-flow motion of Abrikosov vortices at microwave
frequencies, and will therefore exhibit much poorer reflectivities in the
microwave region.

What does the foregoing analysis imply about the ability of a superconducting
film to reflect a GR microwave? In order to answer this question we must
determine the magnitude of the kinetic inductance length scale in the GR case.
First, however, we will take a moment to motivate the idea of the
\textquotedblleft characteristic gravitational impedance of free
space\textquotedblright\ and to consider why objects made of normal matter are
such poor reflectors of GR waves.

\section{The gravitational characteristic impedance of free space}

Wald \cite[Section 4.4]{Wald} has introduced an approximation scheme that
leads to a useful Maxwell-like representation of the Einstein equations of
general relativity. The resulting equations describe the coupling of weak GR
fields to slowly moving matter. In the asymptotically flat spacetime
coordinate system of a distant inertial observer, the four equations in SI
units are
\begin{subequations}
\label{Maxwell-like-eqs}%
\begin{gather}
\mathbf{\nabla\cdot E}_{\text{G}}=-\frac{\rho_{\text{G}}}{\varepsilon
_{\text{G}}}\label{Maxwell-like-eq-1}\\
\mathbf{\nabla\times E}_{\text{G}}=-\frac{\partial\mathbf{B}_{\text{G}}%
}{\partial t}\label{Maxwell-like-eq-2}\\
\mathbf{\nabla\cdot B}_{\text{G}}=0\label{Maxwell-like-eq-3}\\
\mathbf{\nabla\times B}_{\text{G}}=\mu_{\text{G}}\left(  -\mathbf{j}%
_{\text{G}}+\varepsilon_{\text{G}}\frac{\partial\mathbf{E}_{\text{G}}%
}{\partial t}\right)  \label{Maxwell-like-eq-4}%
\end{gather}
where the gravitational analog of the electric permittivity of free space is
given by%
\end{subequations}
\begin{equation}
\varepsilon_{\text{G}}=\frac{1}{4\pi G}=1.2\times10^{9}\text{ SI units}
\label{epsilon_G}%
\end{equation}
and the gravitational analog of the magnetic permeability of free space is
given by%
\begin{equation}
\mu_{\text{G}}=\frac{4\pi G}{c^{2}}=9.3\times10^{-27}\text{ SI units.}
\label{mu_G}%
\end{equation}
The value of $\varepsilon_{\text{G}}$ is fixed by demanding that Newton's law
of gravitation be recovered from the Gauss-like law (\ref{Maxwell-like-eq-1}),
whereas the value of $\mu_{\text{G}}$ is fixed by the linearization procedure
from Einstein's field equations. These two constants express the strengths of
the coupling between sources (i.e., of masses and mass currents, respectively)
and gravitational fields, and are analogous to the two constants
$\varepsilon_{0}$ (the permittivity of free space) and $\mu_{0}$ (the
permeability of free space), which express the strengths of coupling between
sources (charges and charge currents, respectively) and electromagnetic fields
in Maxwell's theory.

In the above set of equations, the field $\mathbf{E}_{\text{G}}$ is the
\emph{gravito}-electric field, which is to be identified with the local
acceleration $\mathbf{g}$ of a test particle produced by the mass density
$\rho_{\text{G}}$, in the Newtonian limit of general relativity. The field
$\mathbf{B}_{\text{G}}$ is the \emph{gravito}-magnetic field produced by the
mass current density $\mathbf{j}_{\text{G}}$ and by the gravitational analog
of the Maxwell displacement current density $\varepsilon_{\text{G}}%
\partial\mathbf{E}_{\text{G}}/\partial t$ of the Ampere-like law
(\ref{Maxwell-like-eq-4}). The resulting magnetic-like field $\mathbf{B}%
_{\text{G}}$ can be regarded as a generalization of the Lense-Thirring field
of general relativity. Because these equations are linear, all fields will
obey the superposition principle not only outside the source (i.e., in the
vacuum), but also within the matter inside the source, provided the field
strengths are sufficiently weak and the matter is sufficiently slowly moving.
Note that the fields $\mathbf{E}_{\text{G}}$ and $\mathbf{B}_{\text{G}}$\ in
the above Maxwell-like equations will be treated as \emph{classical} fields,
just like the fields $\mathbf{E}$ and $\mathbf{B}$ in the classical Maxwell's equations.

As noted earlier, Cooper pairs cannot freely fall along with the ionic lattice
in response to an incident GR wave because the UP forbids such pairs from
having classical trajectories, i.e., from traveling along geodesics. An
incident field $\mathbf{E}_{\text{G}}$ will therefore cause the Cooper pairs
to undergo \emph{non-geodesic} motion, in contrast to the \emph{geodesic
}motion of the ions inside the lattice. This entails the existence of mass
currents (as well as charge currents) from the perspective of a local, freely
falling observer who is located near the surface of the superconducting film
anywhere other than at its center of mass. These\ mass currents will be
describable by a gravitational version of Ohm's law%
\begin{equation}
\mathbf{j}_{\text{G}}(\omega)\mathbf{=\sigma}_{\text{s,G}}(\omega
)\mathbf{E}_{\text{G-inside}}(\omega)\text{ ,} \label{Gravitational-Ohm's-law}%
\end{equation}
where $\mathbf{j}_{\text{G}}(\omega)$ is the mass-current density at frequency
$\omega$, $\sigma_{\text{s,G}}(\omega)=\sigma_{\text{1s,G}}(\omega
)+i\sigma_{\text{2s,G}}(\omega)$ is the complex mass-current conductivity of
the film at the frequency $\omega$ in its linear response to the fields of the
incident GR wave, and $\mathbf{E}_{\text{G-inside}}(\omega)$ is the driving
gravito-electric field inside the film at frequency $\omega$. The existence of
these mass currents can also be inferred from DeWitt's minimal coupling rule
for superconductors (\cite{DeWitt}; see Section 7 below). The real part of the
mass conductivity, $\sigma_{\text{1s,G}}(\omega)$, describes the
superconductor's dissipative response to the incident gravito-electric field,
while the imaginary part, $\sigma_{\text{2s,G}}(\omega)$, describes its
non-dissipative response to the same field. The basic assumption behind
(\ref{Gravitational-Ohm's-law}) is that the mass-current density in any
superconductor responds $linearly$ to a weak GR wave at the driving frequency
\cite{grav-shielding-not-possible}. One should view $\sigma_{\text{s,G}}$\ as
a phenomenological quantity, which, like the electrical conductivity
$\sigma_{\text{s}}$, must be experimentally determined. In any case, the
resulting optics for weak GR waves will be \textit{linear}, just like the
linear optics for weak EM waves.

An important physical property follows from the above Maxwell-like equations,
namely, the characteristic gravitational impedance of free space $Z_{\text{G}%
}$ \cite{Chiao-Wheeler,Kiefer-Weber,Chiao2004}:%
\begin{equation}
Z_{\text{G}}=\sqrt{\frac{\mu_{\text{G}}}{\varepsilon_{\text{G}}}}=\frac{4\pi
G}{c}=2.8\times10^{-18}\text{ SI units.} \label{Z_G}%
\end{equation}
This quantity is a characteristic of the vacuum, i.e., it is a property of
spacetime itself, and it is independent of any of the properties of matter
\textit{per se}. As with $Z_{0}=\sqrt{\mu_{0}/\varepsilon_{0}}=377$ ohms in
the EM case, $Z_{\text{G}}=\sqrt{\mu_{\text{G}}/\varepsilon_{\text{G}}%
}=2.8\times10^{-18}$ SI units will play a central role in all GR radiation
coupling problems. In practice, the impedance of a material object must be
much smaller than this extremely small quantity before any significant portion
of the incident GR-wave power can be reflected. In other words, conditions
must be highly unfavorable for dissipation into heat. Because all classical
material objects have extremely high levels of dissipation compared to
$Z_{\text{G}}$, even at very low temperatures, they are inevitably very poor
reflectors of GR waves \cite{Weinberg,Chiao2004}. The question of
GR-wave reflection from macroscopically coherent quantum systems such as
superconductors requires a separate analysis due to the effectively zero
resistance associated with superconductors, i.e., the dissipationlessness
exhibited by matter in this unique state, at temperatures near absolute zero.

\section{A criterion for the specular reflection of GR waves from
superconducting films}

In the case of EM waves considered above in Section 4, the BCS framework led
us to two related expressions for the behavior of a superconducting thin
film, one for its EM reflectivity (\ref{EM-reflectivity-imped-ratio}) and one
for its EM roll-off frequency (\ref{Impedance-condition}). Now, on the basis
of the similarity of the Maxwell and the Maxwell-like equations, the
identicality of the boundary conditions that follow from these equations, and
the linearity of weak GR-wave optics that follows from the gravitational
version of Ohm's law for superconductors (\ref{Gravitational-Ohm's-law}), we
are led to the following two expressions for the reflectivity and the roll-off
frequency in the GR sector, which are analogous to
(\ref{EM-reflectivity-imped-ratio}) and (\ref{Impedance-condition}),
respectively:%
\begin{subequations}
\begin{align}
{\mathcal{R}}_{\text{G}}  &  =\left\{  1+\left(  2\frac{X_{\text{L,G}}%
}{Z_{\text{G}}}\right)  ^{2}\right\}  ^{-1}\label{GR-reflectivity}\\
\omega_{\text{r,G}}  &  =\pm\frac{Z_{\text{G}}}{2L_{\text{G}}}\text{ .}
\label{GR-refl-cond}%
\end{align}
Once again, we exclude the negative solution in the expression for the upper
roll-off frequency given in (\ref{GR-refl-cond}) as being unphysical.

Pausing for a moment to consider the \emph{lower }roll-off frequency, we find
that a new constraint appears. The H-C effect, which is ultimately responsible
for the mirror-like behavior of the film in the GR case, can only be presumed
to operate when%
\end{subequations}
\begin{equation}
\omega\geq\frac{2\pi v_{\text{s}}}{a}\text{ ,}%
\end{equation}
where $\omega$ is the frequency of the incident wave, $v_{s}$ is the speed of
sound in the medium, and $a$ is the transverse size of a square film. The
physical significance of this constraint becomes apparent when we rewrite it
as%
\begin{equation}
a\geq\frac{2\pi v_{\text{s}}}{\omega}\text{ .}%
\end{equation}
This form of the inequality follows from the fact that neighboring ions
separated by a distance less than $\left(  2\pi v_{\text{s}}\right)  /\omega$
will be mechanically coupled to one another, since there will be sufficient
time for a mechanical signal to propagate from one to the other. Only ions
separated by distances greater than $\left(  2\pi v_{\text{s}}\right)
/\omega$ can be legitimately regarded as separately undergoing free fall in
the presence of a GR wave. Ultimately, however, this additional constraint is
preempted by the inequality already introduced in Section 2,%
\begin{equation}
a\geq\frac{2\pi c}{\omega}\text{ ,}%
\end{equation}
where $\left(  2\pi c\right)  /\omega$ is the wavelength of the incident wave,
since this more stringent requirement must be met for the film to function
as a mirror at all.

Returning to the expression for the \emph{upper} roll-off frequency given in
(\ref{GR-refl-cond}), is it conceivable that this expression could yield a
non-negligible $\omega_{\text{r,G}}$ in the case of a superconducting film? We
begin by noting that the gravitational impedance of free space $Z_{\text{G}}$
can be expressed as%
\begin{equation}
Z_{\text{G}}=\mu_{\text{G}}c\text{ .} \label{Z_G-mu-c}%
\end{equation}
In light of (\ref{GR-refl-cond}) and the smallness of $\mu_{\text{G}}$, as
indicated earlier in (\ref{mu_G}), it would seem highly unlikely that a
superconductor's GR inductance would be small enough to produce a
non-negligible roll-off frequency. Any attempt to construct laboratory-scale
mirrors for GR waves would appear to be doomed from the start. However, as
with $L$ for a thin film in the electromagnetic case, $L_{\text{G }}$ must be
expressible as the product of the permeability and a length scale. In the GR
case, we must use the \emph{gravitational }version of each
parameter. We will neglect the contribution of the gravito-magnetic inductance
$L_{\text{m,G}}$ to the overall gravitational inductance $L_{\text{G}}$ on the
grounds that it will be much smaller than the gravito-kinetic inductance
$L_{\text{k,G}}$ (again, see Appendix A), so that%
\begin{equation}
L_{\text{G}}\approx L_{\text{k,G}}=\mu_{\text{G}}l_{\text{k,G}}\text{ }.
\label{L_G}%
\end{equation}
Inserting (\ref{Z_G-mu-c}) and (\ref{L_G}) into (\ref{GR-refl-cond}), we see
that the permeability cancels out of the numerator and denominator as before,
so that $\omega_{\text{r,G}}$ depends only on the ratio of the speed of light
$c$ to a single parameter -- in this case, the \emph{gravitational }kinetic
inductance length scale $l_{\text{k,G}}$:%
\begin{equation}
\omega_{\text{r,G}}=\frac{\mu_{\text{G}}c}{2\mu_{\text{G}}l_{\text{k,G}}%
}=\frac{c}{2l_{\text{k,G}}}\text{ .} \label{roll-off-freq-BCS-GR}%
\end{equation}

In the electromagnetic case, $l_{\text{k}}$ was given by%
\begin{equation}
l_{\text{k}}=\xi_{0}\left(  \frac{\delta_{\text{p}}}{d}\right)  ^{2}\text{ ,}%
\end{equation}
where the plasma skin depth $\delta_{\text{p}}$ was given by%
\begin{equation}
\delta_{\text{p}}=\sqrt{\frac{m_{\text{e}}}{\mu_{0}n_{\text{e}}e^{2}}}\text{
}. \label{plas-skin-dep-2}%
\end{equation}
In the present context, the coherence length $\xi_{0}$ and the thickness of
the film $d$ must remain the same, since they are \emph{internal }properties
of the film having nothing to do with the strength of coupling to
\emph{external }radiation fields. By contrast, the plasma skin depth
\emph{would }appear to depend on the strength of coupling to external
radiation fields through the presence of $\mu_{0}$ and $e^{2}$ in the
denominator of (\ref{plas-skin-dep-2}). We therefore need to consider the
magnitude of this parameter in the gravitational sector.

For the moment, let us assume that the coupling of Cooper pairs to a GR wave
depends \emph{solely} on their gravitational mass $2m_{\text{e}}$, i.e., that
their electrical charge $2e$ is irrelevant to the gravitational plasma skin
depth and thus to the gravitational kinetic inductance length scale of a
superconducting film. Ultimately, we will reject this approach, since
the Coulomb interaction between the superconductor's Cooper pairs and the
corresponding holes created in the virtual plasma excitation induced within
the film is crucial for understanding how the
film responds to a GR wave. Nonetheless, it is instructive to ignore all
considerations of charge and to presume, for the moment, that Cooper pairs react
to a GR wave solely on the basis of their mass. In fact, the criterion
presented at the end of this section may well be valid for neutral superfluids
(e.g., superfluid helium or a neutral atomic Bose-Einstein condensate), but we
show in the following section that it must be modified in the case of
superconductors to account for the H-C effect.

To obtain the \textquotedblleft gravitational\textquotedblright\ version of
the plasma skin depth $\delta_{\text{p,G}}$, let us make the following
substitution%
\begin{equation}
\frac{e^{2}}{4\pi\varepsilon_{0}}\rightarrow Gm_{\text{e}}^{2}\text{ }%
\end{equation}
in the expression for the plasma skin depth $\delta_{\text{p}}$
(\ref{plas-skin-dep-2}). Note that for this substitution to be valid, we must
treat the electrons as if they were electrically neutral. The
\textquotedblleft gravitational\textquotedblright\emph{ }kinetic inductance
length scale then becomes%
\begin{equation}
l_{\text{k,G }}=\xi_{0}\left(  \frac{\delta_{\text{p,G }}}{d}\right)
^{2}\text{ ,} \label{l_kG-BCS}%
\end{equation}
where$,$ in this spurious approach, $\delta_{\text{p,G}}$ is given by%
\begin{equation}
\delta_{\text{p,G }}=\sqrt{\frac{1}{\mu_{\text{G}}n_{\text{e}}m_{\text{e}}}%
}\text{ }. \label{plas-skin-dep-GR}%
\end{equation}
Assuming here and in subsequent calculations an estimate of $n=n_{\text{e}%
}/2\approx10^{30}$ m$^{-3}$ for the number density of Cooper pairs, one finds
that $\delta_{\text{p,G }}$ is on the order of $10^{13}$ m, which leads to a
value for $l_{\text{k,G}}$ on the order of 10$^{36}$ m. Inserting this
enormous value for $l_{\text{k,G}}$ into (\ref{roll-off-freq-BCS-GR}) yields a
roll-off frequency $\omega_{\text{r,G}}$ of effectively zero, which of course
undermines any practical possibility of GR-wave reflection.

On the grounds that one must eliminate dissipation into heat for the GR-wave
scattering cross-section to become comparable to a square wavelength, Weinberg
has suggested in his discussion of Weber-style resonant bar detectors that
superfluids might function effectively as mirrors for GR waves \cite{Weinberg}%
. The analysis presented here, however, suggests that neutral superfluids
cannot substantially reflect GR waves because of the electrical neutrality of
their mass carriers. (See Appendix C for a brief account of the relation
between the \textquotedblleft impedance\textquotedblright\ argument of the
previous section and Weinberg's analysis of the dissipation problem.) As we
shall see, the fact that a superconductor's mass carriers are not electrically
neutral utterly changes the dynamics of the interaction.

\section{The specular reflection of GR waves}

In Section 2, we argued that the Uncertainty Principle \emph{delocalizes} a
superconductor's Cooper pairs within the material, so that they must exhibit
\emph{non-geodesic} motion rather than the decoherence-induced \emph{geodesic
}motion exhibited by all localized particles, such as freely floating
\textquotedblleft dust particles\textquotedblright\ or the ions in the lattice
of a superconductor. The non-localizability of the Cooper pairs within a
superconducting film leads to charge supercurrents inside the film, which, by
charge conservation and the accumulation of charge at its edges, must produce
a Coulomb electric field inside the film in a virtual plasma excitation of the
material. As a result, enormous Coulomb forces will be created between the
film's negatively charged Cooper pairs and its corresponding, positively
charged holes.

In the GR case, one might think to replace the Coulomb force with the much
weaker Newtonian gravitational force -- as we did in the previous section --
but this amounts to treating the Cooper pairs and holes as if they were
electrically neutral, which is patently unphysical. There can be no H-C effect
in the case of a neutral superfluid, but the situation is entirely different
for a superconductor. This effect, which can appear inside a superconductor, causes a
superconducting film to respond extremely \textquotedblleft
stiffly\textquotedblright\ to an incident GR wave and leads to hard-wall
boundary conditions for the wave. To put the point differently, the
stiffness\ of a superconducting film in its response to an incoming GR wave is
governed by the strength of the Coulomb interaction between the Cooper pairs
and the corresponding holes, and not by their much weaker gravitational
interaction. This fact is reflected in the appearance of the electromagnetic
plasma frequency in the formulas derived below.

Let us begin our analysis of the magnitude of the H-C effect by examining the
quantum probability current density $\mathbf{j}$. This quantity is more basic
than the charge current density $\mathbf{j}_{\text{e}}=nq\mathbf{v}$\ or the
mass current density $\mathbf{j}_{\text{G}}=nm\mathbf{v}$, since one can
derive $\mathbf{j}$ directly from quantum mechanics. It should be regarded as
the \emph{cause }of the charge and mass currents$,$\textit{ }whereas
$\mathbf{j}_{\text{e}}$ and $\mathbf{j}_{\text{G}}$ should be regarded as the
\emph{effects }of $\mathbf{j}$.

Recall that in non-relativistic quantum mechanics $\mathbf{j}$ is given by%
\begin{equation}
\mathbf{j}=\frac{\hbar}{2mi}(\psi^{\ast}\nabla\psi-\psi\nabla\psi^{\ast
})\text{ }, \label{prob. curr. density j}%
\end{equation}
where $m$ is the mass of the non-relativistic particle whose current is being
calculated (here $m=2m_{e}$) and $\psi$ is the wavefunction of the system
(here the Cooper pair's \textquotedblleft condensate
wavefunction\textquotedblright, or London's \textquotedblleft macroscopic
wavefunction\textquotedblright, or Ginzburg and Landau's \textquotedblleft
complex order parameter\textquotedblright). This quantum mechanical quantity satisfies the continuity equation%
\begin{equation}
\nabla\cdot\mathbf{j+}\frac{\partial\rho}{\partial t}=0\text{ ,}
\label{continuity equation}%
\end{equation}
where $\rho=\psi^{\ast}\psi$ is the quantum probability density of the Cooper
pairs. The meaning of (\ref{continuity equation}) is that probability is conserved.

Now let us adopt DeWitt's minimal coupling rule \cite{DeWitt} and make the
following substitution for the momentum operator:%
\begin{subequations}
\begin{align}
\mathbf{p}  &  \rightarrow\mathbf{p}-q\mathbf{A}-m\mathbf{h}\text{
or}\label{DeWitt}\\
\frac{\hbar}{i}\nabla &  \rightarrow\frac{\hbar}{i}\nabla-q\mathbf{A}%
-m\mathbf{h}\text{ ,}%
\end{align}
where $q=2e$, $m=2m_{e}$, $\mathbf{A}$ is the electromagnetic vector
potential, and $\mathbf{h}$ is DeWitt's gravitational vector potential
\cite{strains} (here and henceforth the dependence on space and time
$(\mathbf{r},t)$ of all field quantities will be suppressed as understood). In
what follows, both $\mathbf{A}$ and $\mathbf{h}$ fields will be treated as
\emph{classical} fields, whereas $\mathbf{j}$ and $\rho$ will be treated as
time-dependent \emph{quantum} operators, in a \emph{semi-classical} treatment
of the interaction of radiation with matter.

We shall also follow DeWitt in adopting the radiation gauge conditions for
both $\mathbf{A}$ and $\mathbf{h}$, namely, that%
\end{subequations}
\begin{equation}
\nabla\cdot\mathbf{A}=0\text{ and }\nabla\cdot\mathbf{h}=0\text{ ,}%
\end{equation}
and that the scalar potentials for both the EM and GR fields vanish
identically everywhere. This choice of gauge means that the coordinate system
being employed is that of an inertial observer located at infinity.

Since it is the case that%
\begin{equation}
\frac{\hbar}{2mi}(\psi^{\ast}\nabla\psi-\psi\nabla\psi^{\ast})=\frac{1}%
{m}\operatorname{Re}\left(  \psi^{\ast}\frac{\hbar}{i}\nabla\psi\right)
\text{ ,} \label{prob current as real part of psi* grad psi}%
\end{equation}
we can apply DeWitt's minimal coupling rule to
(\ref{prob current as real part of psi* grad psi}) to obtain%
\begin{equation}
\mathbf{j}=\frac{1}{m}\operatorname{Re}\left(  \psi^{\ast}\left\{  \frac
{\hbar}{i}\nabla-q\mathbf{A}-m\mathbf{h}\right\}  \psi\right)  \text{ }.
\label{j with A and h}%
\end{equation}
The continuity equation (\ref{continuity equation}) is still satisfied by
(\ref{j with A and h}), provided that one also applies the same minimal
coupling rule to the time-dependent Schr\"{o}dinger equation, in which the
Hamiltonian becomes%
\begin{equation}
H=\frac{\left(  \mathbf{p}-q\mathbf{A}-m\mathbf{h}\right)  ^{2}}{2m}+V\text{
},
\end{equation}
where the first term on the right-hand side represents the kinetic energy
operator, and $V$ is the potential energy operator.

In the special case of \emph{neutral}, classical \textquotedblleft dust
particles\textquotedblright\ in the presence of a GR wave, $q=0$ and thus
$q\mathbf{A}=0$ (as well as $V=0$). The classical Hamilton's function
$H(\mathbf{p},\mathbf{q})$ then becomes%
\begin{equation}
H(\mathbf{p},\mathbf{q})=\frac{\left(  \mathbf{p}-m\mathbf{h}\right)  ^{2}%
}{2m}\text{ }. \label{Neutral DeWitt Hamiltonian}%
\end{equation}
Defining the canonical momentum classically as $\mathbf{p}=m\mathbf{v}_\text{can}$, where
$\mathbf{v}_\text{can}$ is the canonical velocity, it will be the case for neutral, classical dust
particles\ that%
\begin{equation}
H=\frac{1}{2}m\left(  \mathbf{v}_\text{can}-\mathbf{h}\right)  ^{2}=0\text{ \ \ or
\ \ }\mathbf{v}_\text{can}=\mathbf{h}\text{ ,} \label{canonical_velocity}%
\end{equation}
as seen by a distant inertial observer, since a passing GR wave cannot impart
any kinetic energy to noninteracting, freely-falling particles. The dust
particles will be carried along \emph{with }space, which follows directly from
the EP.

On the other hand, when (\ref{Neutral DeWitt Hamiltonian}) is viewed as a
quantum Hamiltonian operator, it implies that neutral,
\emph{quantum-mechanical} particles will acquire a kinetic energy equal to
$\frac{1}{2}m\mathbf{h}^{2}$ when they are in a nonlocalizable, gap-protected,
zero-momentum eigenstate ($\mathbf{p}=\mathbf{0,}$ where $\mathbf{p}$ is the
canonical momentum\textbf{)}. In accord with first-order time-dependent
perturbation theory, such particles must remain in their ground state in the
presence of a GR wave whose frequency is less than the BCS gap frequency. They
will therefore rigidly resist the stretching and squeezing of space caused by
such a wave. In other words, they will be locally accelerated \emph{through
}space, acquiring kinetic energy in the process. In the case of superfluid
helium, for example, in which the basic components of the material are both
electrically neutral and quantum-mechanically protected from excitations by
the roton gap, mass supercurrents will be created that carry kinetic energy
extracted from the wave.

Now let us consider the case of a type I superconductor. Before the arrival of
a GR wave, the superconductor's Cooper pairs will be in a zero-momentum
eigenstate:%
\begin{equation}
\psi=C\exp(i\frac{\mathbf{p}_{0}\cdot\mathbf{r}}{\hbar})\text{ where
}\mathbf{p}_{0}=\mathbf{0}\text{ .}\label{plane wave state}%
\end{equation}
Again, in accord with first-order time-dependent perturbation theory, this
initial wavefunction must remain unchanged to lowest order by the radiative
perturbations arising from either $\mathbf{A}$ or $\mathbf{h}$ after the
arrival of a wave whose frequency is less than the BCS gap frequency of the
material. If one evaluates (\ref{j with A and h}) using the unperturbed state
(\ref{plane wave state}), one finds that%
\begin{align}
\mathbf{j} &  =\frac{1}{m}\operatorname{Re}\left(  \psi^{\ast}\left\{
\frac{\hbar}{i}\nabla-q\mathbf{A}-m\mathbf{h}\right\}  \psi\right)
\label{j with A and h simplified}\\
&  =\frac{1}{m}\left(  -q\mathbf{A}-m\mathbf{h}\right)  \psi^{\ast}\psi\text{
}.\nonumber
\end{align}
From this one can define the \textquotedblleft quantum velocity
field\textquotedblright\ $\mathbf{v}$,%
\begin{equation}
\mathbf{v}=\frac{\mathbf{j}}{\rho}=\frac{\mathbf{j}}{\psi^{\ast}\psi}\text{ ,}%
\end{equation}
whose local expectation value is the local group velocity of a Cooper pair
\cite{group velocity}. It thus follows that%
\begin{equation}
\mathbf{v}=-\frac{q}{m}\mathbf{A-h}\label{velocity field}%
\end{equation}
inside a superconducting film after the arrival of a GR wave. This velocity is
the \emph{kinetic} velocity of the quantum supercurrent, and not the canonical velocity of a classical dust particle given in (\ref{canonical_velocity}), in the sense that
$\frac{1}{2}m\mathbf{v}^{2}$ is the local kinetic energy of the quantum supercurrent.

The generation of mass supercurrents inside a superconductor by the GR wave
will also produce \emph{charge }supercurrents inside the superconductor, since
$q$ is not zero for Cooper pairs. These supercurrents will electrically
polarize the superconductor, which will set up an \emph{internal }$\mathbf{A}$
field -- even in the absence of any incident EM wave. Thus, the term
$(-q/m)\mathbf{A}$ on the right-hand side of (\ref{velocity field}) will
\emph{not} be zero inside a superconductor in the presence of a GR wave.
Herein lies the possibility of mirror-like reflection of GR waves from
superconducting thin films.

Taking the partial derivative of (\ref{velocity field}) with respect to time,
and defining the meaning of this derivative in the sense of Heisenberg's
equation of motion for the kinetic velocity operator $\mathbf{v}$, one obtains
an operator equation of motion that has the same form as Newton's 2nd law of
motion, namely,%
\begin{equation}
m\frac{\partial}{\partial t}\mathbf{v=}m\frac{\partial^{2}}{\partial t^{2}%
}\mathbf{x}=m\mathbf{a}=q\mathbf{E}+m\mathbf{E}_{\text{G}}\text{
,}\label{Newton-like EOM}%
\end{equation}
where, by our gauge choice, $\mathbf{E}$\ and $\mathbf{E}_{\text{G}}$ inside
the superconductor are related to the vector potentials $\mathbf{A}$ and
$\mathbf{h}$, respectively, by%
\begin{equation}
\mathbf{E}=-\frac{\partial}{\partial t}\mathbf{A}\text{ and }\mathbf{E}%
_{\text{G}}=-\frac{\partial}{\partial t}\mathbf{h}\text{ .}%
\label{E_to_A_and_E_G_to_h}%
\end{equation}
Both $\mathbf{E}$\ and $\mathbf{E}_{\text{G}}$ will be treated here as
classical fields. Following the presentation in Section 5, $\mathbf{E}%
_{\text{G}}$ is the gravito-electric field that appears in the Maxwell-like
equations, which is equivalent to the acceleration $\mathbf{g}$ of a local, classical
test particle due to gravity, in accord with the EP. The physical
interpretation of the Newton-like equation of motion (\ref{Newton-like EOM})
is that the internal $\mathbf{E}$ and $\mathbf{E}_{\text{G}}$ fields act upon
the charge $q$ and the mass $m$, respectively, of the Cooper pairs, to produce
an acceleration field $\mathbf{a}$ of these pairs (in the sense of Ehrenfest's
theorem) inside a superconducting film.

For all fields that vary sinusoidally with the same exponential phase factor
$\exp(-i\omega t),$ (\ref{Newton-like EOM}) leads to the following
linear-response equation at the frequency $\omega$:%
\begin{equation}
\mathbf{x}=-\frac{1}{\omega^{2}}\left(  \frac{q}{m}\mathbf{E}+\mathbf{E}%
_{\text{G}}\right)  \text{ }\mathbf{.} \label{lin-resp x by E and E_G}%
\end{equation}
The mass current density source term in the Ampere-like law
(\ref{Maxwell-like-eq-4}) of the Maxwell-like equations is then given by%
\begin{align}
\mathbf{j}_{\text{G}}  &  =nm\mathbf{v}=nm\frac{\partial}{\partial
t}\mathbf{x}\label{j_G vs. E and E_G}\\
&  =nm(-i\omega)\mathbf{x}\nonumber\\
&  =i\frac{n}{\omega}\left(  q\mathbf{E+}m\mathbf{E}_{\text{G}}\right)  \text{
}.\nonumber
\end{align}
The total force acting on a given Cooper pair under such circumstances is thus
\cite{Schiff-Barnhill}%
\begin{equation}
\mathbf{F}_{\text{tot}}=q\mathbf{E}+m\mathbf{E}_{\text{G}}\text{ ,}
\label{F as sum of E and E_G}%
\end{equation}
which is to say that $\mathbf{F}_{\text{tot}}$ depends on a \emph{linear
combination }of the internal $\mathbf{E}$ and $\mathbf{E}_{\text{G}}$ fields,
or, equivalently, that a superconductor will respond \emph{linearly }to a
sufficiently weak incident GR wave.

When a superconductor is operating in its \emph{linear response regime }in the
presence of a weak incident GR wave, the following direct proportionalities
will hold:%
\begin{equation}
\mathbf{F}_{\text{tot}}\propto\mathbf{E\propto\mathbf{E}_{\text{G}}}\text{
}\mathbf{.} \label{F proportional to E and g}%
\end{equation}
Let us therefore define a proportionality constant $\Xi,$ such that%
\begin{equation}
\mathbf{F}_{\text{tot}}=\Xi q\mathbf{E}\text{ .} \label{F_prop_to_qE}%
\end{equation}
We shall call this dimensionless proportionality constant the
\textquotedblleft fractional correction factor\textquotedblright\ of the total
force acting upon a given Cooper pair, relative to a purely electrical force
acting on the same pair.

At this point, it would be customary to ignore the extremely weak
gravitational forces generated internally within the superconducting film.
That is to say, one would normally set the gravitational field $\mathbf{E}%
_{\text{G}}$ inside the film identically equal to zero everywhere by declaring
that $\Xi=1$, exactly. One could then solve the essentially
\emph{electromagnetic\ }problem of virtual plasma excitations produced inside
the film in its linear response to a weak incident EM or GR wave.

But this simplification will not suffice in the present context, since we want
to understand the dynamics of the system when one takes into account the
\emph{combined }effect of the internal electric field $\mathbf{E}$ and
internal gravito-electric field $\mathbf{E}_{\text{G}}$, both of which will be
produced in association with the electrical polarization of the superconductor
induced by an incident EM or GR wave. Although the impact on the
\emph{electro}dynamics of the system will be negligible, the impact on its
\emph{gravito}-electrodynamics will be enormous. Let us then use
(\ref{F as sum of E and E_G}) and (\ref{F_prop_to_qE}) to express the
relationship between the $\mathbf{E}$ and $\mathbf{E}_{\text{G}}$ fields
inside a superconducting film when $\Xi\neq1,$ i.e., when the gravitational
forces within the film, however tiny, are explicitly taken into account:%
\begin{equation}
\mathbf{E}=\frac{1}{\Xi-1}\frac{m}{q}\mathbf{E}_{\text{G}}\text{ }.
\label{E_in_terms_of_E_G}%
\end{equation}
Substituting this expression into (\ref{j_G vs. E and E_G}), we obtain
\cite{critical-field}%
\begin{equation}
\mathbf{j}_{\text{G}}=i\frac{\Xi}{\Xi-1}\frac{nm}{\omega}\mathbf{E}_{\text{G}%
}\text{ }, \label{j_G_in_terms_of_Ksi_and_E_G}%
\end{equation}
from which it follows that the mass conductivity of the film $\sigma
_{\text{G}}$ is given by%
\begin{equation}
\sigma_{\text{G}}=i\left(  \frac{\Xi}{\Xi-1}\right)  \frac{nm}{\omega}%
\propto\frac{1}{\omega}\text{ ,} \label{sigma_G in terms of Ksi}%
\end{equation}
implying an inductive response to internal fields on the part of the mass
currents $\mathbf{j}_{\text{G}}$ within the film. Note that $\sigma_{\text{G}%
}$ can in principle become extremely large when $\Xi\rightarrow1$, and
therefore that $\mathbf{j}_{\text{G}}$\ can become extremely large.

Let us consider first the effect of the gravitational force between the Cooper
pairs and holes on the plasma frequency. We start from (\ref{Newton-like EOM})
in the form%
\begin{equation}
m\frac{\partial^{2}}{\partial t^{2}}\mathbf{x}=-m\omega^{2}\mathbf{x}%
=q\mathbf{E}+m\mathbf{E}_{\text{G}}\mathbf{=}\Xi q\mathbf{E}\text{ ,}%
\end{equation}
so that%
\begin{equation}
\mathbf{x=-}\Xi\frac{q}{m\omega^{2}}\mathbf{E}\text{ .}%
\end{equation}
The electric polarization of the superconductor will then be%
\begin{equation}
\mathbf{P}=nq\mathbf{x}=-\Xi\frac{nq^{2}}{m\omega^{2}}\mathbf{E}%
=\chi_{\text{p}}^{\prime}\varepsilon_{0}\mathbf{E}\text{ ,}%
\end{equation}
where $\chi_{\text{p}}^{\prime}$ is the modified plasma susceptibility. Since
this susceptibility can be expressed as%
\begin{equation}
\chi_{\text{p}}^{\prime}=-\frac{\omega_{\text{p}}^{\prime2}}{\omega^{2}}\text{
,}%
\end{equation}
it follows that the square of the modified plasma frequency is given by%
\begin{equation}
\omega_{\text{p}}^{\prime2}=\Xi\frac{nq^{2}}{m\varepsilon_{0}}\text{ .}
\label{modified plasma frequency in terms of Ksi}%
\end{equation}
We thus expect that the fractional correction factor $\Xi$, which takes into
account the gravitational forces between the Cooper pairs and holes, will lead
to an extremely small correction to the standard formula for the plasma frequency.

To determine the magnitude of $\Xi,$ we begin with the quantum form of
Newton's second law (\ref{Newton-like EOM}), rewritten as%
\begin{equation}
\frac{\partial}{\partial t}\mathbf{v}=\frac{q}{m}\mathbf{E}+\mathbf{E}%
_{\text{G}}\ . \label{Newton-2nd-law/m}%
\end{equation}
Multiplying both sides by $nq$, one obtains a current-density form of the same
equation:%
\begin{equation}
\frac{\partial(nq\mathbf{v)}}{\partial t}=\frac{\partial}{\partial
t}\mathbf{j}_{\text{e}}=\frac{nq^{2}\mathbf{E}}{m}+nq\mathbf{E}_{\text{G}}\ .
\label{Current EOM}%
\end{equation}
Let us evaluate all quantities in this equation at a point $P$ along the edge
of the superconducting film where the ionic lattice abruptly ends and the
vacuum begins:%
\begin{equation}
\left.  \frac{\partial}{\partial t}\mathbf{j}_{\text{e}}\right\vert
_{P}=\left.  \frac{nq^{2}\mathbf{E}}{m}\right\vert _{P}+\left.  nq\mathbf{E}%
_{\text{G}}\right\vert _{P}\ . \label{Current EOM evaluated at P}%
\end{equation}
We will assume that the incident radiation fields that excite the Cooper-pair
plasma are tightly focused onto a diffraction-limited Gaussian-beam spot size
located at the center of the square film. We will also assume that the
radiative excitation is impulsive in nature, so that the plasma can oscillate
freely after the radiation is abruptly turned off. Thus the point $P$ at the
edge of the film at which all quantities in (\ref{Current EOM evaluated at P})
are to be evaluated, is far away from the center of the film, where the
incident radiation fields can impulsively excite the film into free plasma oscillations.

Taking the divergence of both sides of (\ref{Current EOM evaluated at P}), we
obtain at point $P$%
\begin{equation}
\left.  \frac{\partial}{\partial t}\left(  \nabla\cdot\mathbf{j}_{\text{e}%
}\right)  \right\vert _{P}=\left.  \frac{nq^{2}}{m}\left(  \nabla
\cdot\mathbf{E}\right)  \right\vert _{P}+\left.  nq\left(  \nabla
\cdot\mathbf{E}_{\text{G}}\right)  \right\vert _{P}.
\label{Time derivate of divergence of j_e}%
\end{equation}
But with the help of the continuity equation
\begin{equation}
\nabla\cdot\mathbf{j}_{\text{e}}+\frac{\partial}{\partial t}\rho_{\text{e}}=0
\end{equation}
and the 1st Maxwell and 1st Maxwell-like equations%
\begin{equation}
\nabla\cdot\mathbf{E=}\frac{\rho_{\text{e}}}{\varepsilon_{0}}\text{ and
}\nabla\cdot\mathbf{E}_{\text{G}}\mathbf{=-}\frac{\rho_{\text{G}}}%
{\varepsilon_{\text{G}}}\text{ ,} \label{1st Maxwell and Maxwell-like Eqs.}%
\end{equation}
we can rewrite (\ref{Time derivate of divergence of j_e}) as a differential
equation for the charge and mass densities at point $P$
\cite{density-operator-EOM}:%
\begin{equation}
-\frac{\partial^{2}}{\partial t^{2}}\rho_{\text{e}}=\frac{nq^{2}}%
{m\varepsilon_{0}}\rho_{\text{e}}-\frac{nq}{\varepsilon_{\text{G}}}%
\rho_{\text{G}}. \label{EOM for rho_e and rho_G}%
\end{equation}
These densities will oscillate freely in time at point $P$ at the edge of the
film, where both charge and mass can accumulate, after the impulsive
excitation at the center of the film has been turned off. We then use the fact
that the accumulated Cooper-pair mass density at point $P$ must be related to
the accumulated Cooper-pair charge density at point $P$ by%
\begin{equation}
\rho_{\text{G}}=\frac{m}{q}\rho_{\text{e}}\text{ ,}%
\end{equation}
since each Cooper pair accumulating at the edge of the film carries with it
both a charge $q$ and a mass $m$. Then at point $P$
(\ref{EOM for rho_e and rho_G}) becomes%
\begin{equation}
-\frac{\partial^{2}}{\partial t^{2}}\rho_{\text{e}}=\frac{nq^{2}}%
{m\varepsilon_{0}}\rho_{\text{e}}-\frac{nm}{\varepsilon_{\text{G}}}%
\rho_{\text{e}}\text{ ,}%
\end{equation}
which leads to the simple harmonic equation of motion%
\begin{equation}
\frac{\partial^{2}}{\partial t^{2}}\rho_{\text{e}}+\frac{nq^{2}}%
{m\varepsilon_{0}}\rho_{\text{e}}-\frac{nm}{\varepsilon_{\text{G}}}%
\rho_{\text{e}}=\frac{\partial^{2}}{\partial t^{2}}\rho_{\text{e}}%
+\omega_{\text{p}}^{\prime2}\rho_{\text{e}}=0\text{ ,}%
\end{equation}
where the square of the modified plasma frequency $\omega_{\text{p}}^{\prime}$
is given by%
\begin{equation}
\omega_{\text{p}}^{\prime2}=\left(  1-\frac{m^{2}}{q^{2}}\frac{Z_{\text{G}}%
}{Z_{0}}\right)  \frac{nq^{2}}{m\varepsilon_{0}}.
\label{modi_plas_freq_Z_0_Z_G}%
\end{equation}
Here we have made use of the fact that $Z_{0}=\left(  c\varepsilon_{0}\right)
^{-1}$ and that $Z_{\text{G}}=\left(  c\varepsilon_{\text{G}}\right)
^{-1}=4\pi G/c$. Comparing (\ref{modi_plas_freq_Z_0_Z_G}) with
(\ref{modified plasma frequency in terms of Ksi}), we arrive at the following
expression for $\Xi$:%
\begin{align}
\Xi &  =1-\frac{m^{2}}{q^{2}}\frac{Z_{\text{G}}}{Z_{0}}\label{Ksi}\\
&  =1-\frac{4\pi\varepsilon_{0}Gm_{\text{e}}^{2}}{e^{2}}\nonumber\\
&  \approx1-\frac{1}{4.2\times10^{42}}\text{ .}\nonumber
\end{align}
The fractional correction factor $\Xi$ does indeed differ from unity by an
extremely small amount, equal to the reciprocal of the ratio of the
electrostatic force to the gravitational force between two electrons given by
(\ref{Ratio of electrical to gravitational forces}).

The implication of (\ref{Ksi}) for the \emph{electro}dynamics of a
superconductor is that the size of the modified plasma frequency given by
(\ref{modified plasma frequency in terms of Ksi}) will be smaller\emph{ }than
the standard value, albeit by a mere $4$ parts in $10^{42}$. Although this
difference is extremely small, the fact that the modified plasma frequency is
\emph{smaller }rather than \emph{larger }points to a surprising fact: the
Cooper-pair holes created inside a superconducting film by an incident EM or
GR microwave must be gravitationally repelled by, rather than attracted to,
the corresponding Cooper pairs in the film, i.e., the holes must have the
equivalent of negative mass and must therefore behave analogously to buoyant
bubbles inside a fluid in the Earth's gravity. This would be a troubling
result, were it not for the fact that the holes, like bubbles,\ cannot exist
independently in the vacuum. The existence of negative-mass \emph{pseudo}%
-particles (i.e., holes) within the film does \emph{not} imply the possibility
of shielding \emph{static, longitudinal} gravito-electric fields, which
requires the existence of \emph{real} particles with negative mass in the
vacuum. That is to say, the existence of these pseudo-particles does \emph{not
}imply the possibility of anti-gravity devices
\cite{grav-shielding-not-possible}.

The real significance of $\Xi$ lies in its impact on the \emph{gravito-}%
electrodynamics of a superconducting film. In particular, the result given in
(\ref{Ksi}) leads to an enhancement of the film's mass conductivity
$\sigma_{\text{G}}$ by the enormous factor of $4.2\times10^{42}$, which is
what we have been calling the Heisenberg-Coulomb effect. Specifically, the
expression for the mass conductivity given in (\ref{sigma_G in terms of Ksi})
can now be reduced to%
\begin{equation}
\sigma_{\text{G }}=-i\frac{nq^{2}\Xi}{m\omega}\frac{Z_{0}}{Z_{\text{G}}}\text{
,} \label{sigma_G in terms of Ksi_reduced}%
\end{equation}
or, equivalently \cite{sigma at DC},%
\begin{equation}
\sigma_{1,\text{G}}=0\text{ \ \ and \ \ }\sigma_{2,\text{G}}=-\frac{nq^{2}\Xi
}{m\omega}\frac{Z_{0}}{Z_{\text{G}}}\text{ .} \label{sigma_G_Ksi_reduced}%
\end{equation}
Let us use this result to calculate the GR reflectivity of a superconducting film.

Recall the relationship given earlier in (\ref{inductive-reactance}) between
the inductance of the film and its nondissipative conductivity. Let us assume,
once again, that the gravito-magnetic inductance is negligible when compared
to the gravitational kinetic inductance (which is justified in Appendix A). We
can then equate the gravitational inductance of the film $L_{\text{G}}$ with
$L_{\text{k,G}}$ and use (\ref{plas-skin-dep-2}), (\ref{plas-skin-dep-GR}),
(\ref{sigma_G_Ksi_reduced}) to express the latter as%
\begin{align}
L_{\text{k,G }}  &  =\frac{1}{\omega\sigma_{2,\text{G}}d}=-\frac{m}{nq^{2}\Xi
}\frac{1}{d}\frac{Z_{\text{G}}}{Z_{0}}\\
&  \approx-\mu_{\text{G}}d\left(  \frac{\delta_{\text{p,G}}}{d}\right)
^{2}\frac{m^{2}Z_{\text{G}}}{q^{2}Z_{0}}\nonumber\\
&  =-\mu_{\text{G}}d\left(  \frac{\delta_{\text{p}}}{d}\right)  ^{2}%
\nonumber\\
&  =-\mu_{\text{G}}l_{\text{k,G}}^{^{\prime}}\text{ ,}\nonumber
\end{align}
where the \emph{corrected }gravitational kinetic inductance length scale
$l_{\text{k,G}}^{\prime}$ is given by%
\begin{equation}
l_{\text{k,G}}^{\prime}=d\left(  \frac{\delta_{\text{p}}}{d}\right)
^{2}\text{ .} \label{l_kG-plasma-corr-factor}%
\end{equation}
But this is just the EM kinetic inductance length scale $l_{\text{k,p}}$ that
appears in the collisionless plasma model presented in Appendix B. Notice that
this expression differs from the BCS expression given in (\ref{l_k}) in
Section 4 by a factor on the order of unity, i.e., $d/\xi_{0}$, which is due
to the fact that the plasma model knows nothing of the BCS coherence length
scale. Nonetheless, the appearance of $\delta_{\text{p}}$ in
(\ref{l_kG-plasma-corr-factor}) highlights the importance of plasma concepts
for correcting the approach adopted at the end of Section 6. The H-C effect
reduces the GR kinetic inductance length scale $l_{\text{k,G}}$ by $42$ orders
of magnitude, to the level of the EM kinetic inductance length scale
$l_{\text{k,p}}$ ($\approx l_{\text{k}}$), thereby increasing the magnitude of
the GR roll-off frequency $\omega_{\text{r,G}}$ by the same factor, to the
level of the EM roll-off frequency $\omega_{\text{r}}$.

Two possible criticisms of this analysis immediately come to mind. First, the
group velocity of a Cooper pair given by (\ref{velocity field}) is predicted
to be superluminal, even for extremely small values of the dimensionless
strain $h_{+}$ of an incident GR wave \cite{strains}. Using
(\ref{E_to_A_and_E_G_to_h}), (\ref{j_G vs. E and E_G}), and
(\ref{j_G_in_terms_of_Ksi_and_E_G}) to solve for $\left\vert \mathbf{v}%
/c\right\vert $, one finds that%
\begin{equation}
\left\vert \frac{\mathbf{v}}{c}\right\vert =\frac{1}{c}\frac{\Xi}{\Xi
-1}\left\vert \mathbf{h}\right\vert =\frac{1}{2}\frac{\Xi}{\Xi-1}\left\vert
h_{+}\right\vert \text{ .}%
\end{equation}
Even for an arbitrarily chosen, extremely small value of $\left\vert
h_{+}\right\vert \approx10^{-40}$ (which, for a 6 GHz GR wave, corresponds to
an incident power flux on the order of $10^{-16}$ W m$^{-2}$), the value given
in (\ref{Ksi}) leads to a velocity roughly one hundred times the speed of
light. This apparent violation of special relativity suggests that the
response of a superconductor to a GR-wave field will in general be nonlinear,
invalidating our assumption of linearity in (\ref{F proportional to E and g}%
)\textbf{.}

However, group velocities much larger than $c$ (infinite, even) have been
experimentally demonstrated \cite{Chiao-Steinberg}. In particular, photon
tunneling-time measurements confirm the \textquotedblleft
Wigner\textquotedblright\ transfer time, which is a measure of an effective
group velocity broadly applicable to quantum scattering processes.
Wigner's analysis \cite{Wigner} assumes a \emph{linear }relation between the
initial and final states of a quantum system, and yields a transfer time that
is proportional to the derivative of the phase of the system's transfer
function with respect to the energy of the incident particle. In the present
context, this implies that the Wigner time will be zero, since the phase of
the Cooper-pair condensate remains constant everywhere, and stays unchanged
with time and energy, due to first-order time-dependent perturbation theory
(i.e., assuming that no pair-breaking or any other quantum excitation is
allowed \cite{pair-breaking}). Returning to Figure \ref{fig:ion-trajectories},
the Wigner time implies that an observer located at the center of mass of the
superconductor who spots a Cooper pair at point B during the passage of the
wave will see the pair disappear and then \emph{instantaneously }re-appear at
point A. This kind of \emph{simultaneity} (as seen by the observer at the
center of mass of the system)\ is a remarkable consequence of quantum theory,
but it does not violate special relativity, nor does it invalidate the
assumption of linearity.

We have already touched on the second criticism, namely, that the analysis
presented here is defective because it does not register the BCS gap
frequency. In particular, ohmic dissipation will occur at frequencies above
the material's BCS gap frequency \cite{Tinkham} and will damp out the free
plasma oscillations that are otherwise predicted to occur in
(\ref{modi_plas_freq_Z_0_Z_G}). In response, we note that these
\emph{dissipative }effects cannot\emph{ }alter the ratio given by
(\ref{Ratio of electrical to gravitational forces}) that appears in the
\emph{nondissipative }factor $\Xi$ given in (\ref{Ksi}). Fundamentally, it is
the strength of the Coulomb force, and not the strength of the gravitational
force, that dictates the strength of a superconducting film's response to an
incident GR wave.

\section{The negligibility of single-bounce transduction}

It is important to address the concern that an incoming GR wave will be
partially or completely transduced into an outgoing EM wave by a
superconducting film instead of being specularly reflected by the film. Recall
that the Cooper pairs within the film \emph{cannot} undergo free fall along
with its lattice in the presence of an incident GR wave, contrary to a naive
application of the EP to all particles. Instead, Cooper pairs must undergo
\emph{non-geodesic} motion, in contrast to the \emph{geodesic }motion of the
ions in the film's lattice. This leads to a non-zero quantum current density,
one that carries mass and charge. Therefore, time-varying mass currents and
time-varying charge currents will be generated by an incident GR wave. The
latter will cause at least some of the incoming GR-wave energy to be
transduced into an outgoing EM wave. More succinctly, the film will behave
like an EM antenna. Appreciable transduction would be an interesting result in
its own right, but it turns out to be negligible. The transduction effect is
necessarily present in the interaction of a superconducting film with a GR
wave, but it does not undermine the film's ability to specularly reflect the wave.

The size of the transduction effect can be determined from a consideration of
the charge supercurrent density generated within a superconducting film by an
incident GR wave. Let us examine the case of a GR plane wave normally incident
upon a superconducting film located at the plane $x=0,$ in the absence of any
incident EM radiation. In this situation, the charge supercurrent generated by
a GR wave will be generated as a \emph{current sheet}. If a GR wave is
incident upon the film only from the left, say, the charge supercurrent
generated in the film will nonetheless radiate EM radiation symmetrically,
i.e., in both the $+x$ and $-x$ directions. This follows from the bilateral
symmetry of the current sheet, which takes the form%
\begin{equation}
\mathbf{j}_{\text{e}}=\mathbf{j}_{0}\delta(x)\exp(-i\omega t)
\end{equation}
around $x=0$ (here and henceforth we suppress the polarization vectors of the
currents and fields because they are all transverse to the $x$ axis). The
current sheet will radiate by coupling, via the Cooper pairs' charge $q=2e,$
to an electric field $\mathbf{E}=-\partial\mathbf{A}/\partial t$ (in the
radiation gauge) and to a magnetic field $\mathbf{B}=\nabla\times\mathbf{A}$.\ 

Having chosen the radiation gauge, in which $\nabla\cdot\mathbf{A}=0$ and in
which the scalar potential is identically zero everywhere, we can begin with
the EM wave equation in terms of $\mathbf{A}$ and $\mathbf{j}_{\text{e}}$:%
\begin{equation}
\nabla^{2}\mathbf{A}-\frac{1}{c^{2}}\frac{\partial^{2}\mathbf{A}}{\partial
t^{2}}=-\mu_{0}\mathbf{j}_{\text{e}}\text{ .} \label{EM-wave-equation}%
\end{equation}
Let us assume once again that all time variations are sinusoidal at an angular
frequency $\omega,$ so that we can make the replacements%
\begin{equation}
\mathbf{A}\rightarrow\mathbf{A}\exp(-i\omega t)\text{ and }\mathbf{j}%
_{\text{e}}\rightarrow\mathbf{j}_{\text{e}}\exp(-i\omega t)\text{ .}%
\end{equation}
Let us also take advantage of the symmetry inherent in the problem, so that we
can reduce (\ref{EM-wave-equation}) to a Helmholtz equation in a single
dimension for the transverse amplitudes $A$ and $j_{\text{e}}$:%
\begin{equation}
\frac{\partial^{2}A}{\partial x^{2}}+k^{2}A=-\mu_{0}j_{\text{e}}=-\mu_{0}%
j_{0}\delta(x)\text{ }. \label{Helmholtz}%
\end{equation}
The delta function in (\ref{Helmholtz}) vanishes everywhere except at the
origin $x=0,$ so that for all $x\neq0$ this equation becomes a 1D homogeneous
Helmholtz equation%
\begin{equation}
\frac{\partial^{2}A}{\partial x^{2}}+k^{2}A=0\text{ .}%
\end{equation}
By the principle of causality and the bilateral symmetry of the film, we can
then restrict the possible solutions of this equation to \textit{outgoing}
plane waves symmetrically emitted from the film, so that%
\begin{subequations}
\begin{align}
A  &  =\alpha\exp(+ikx)\text{ for }x>0\\
A  &  =\alpha\exp(-ikx)\text{ for }x<0
\end{align}
for the same value of $\alpha$, which is determined by the strength of the
delta function as follows:%
\end{subequations}
\begin{gather}
\lim_{\varepsilon\rightarrow0}%
{\displaystyle\int\limits_{-\varepsilon}^{+\varepsilon}}
dx\left(  \frac{\partial^{2}A}{\partial x^{2}}+k^{2}A\right)  =\lim
_{\varepsilon\rightarrow0}\left.  \frac{\partial A}{\partial x}\right\vert
_{-\varepsilon}^{+\varepsilon}\nonumber\\
=-\mu_{0}j_{0}\lim_{\varepsilon\rightarrow0}%
{\displaystyle\int\limits_{-\varepsilon}^{+\varepsilon}}
dx\delta(x)=-\mu_{0}j_{0}\text{ }.\nonumber
\end{gather}
For $\varepsilon>0,$ the derivatives of $A$ are%
\begin{equation}
\lim_{\varepsilon\rightarrow0}\left.  \frac{\partial A}{\partial x}\right\vert
_{{}}^{+\varepsilon}=\lim_{\varepsilon\rightarrow0}\left(  +ik\alpha
\exp(+ik\varepsilon)\right)  =+ik\alpha
\end{equation}
and%
\begin{equation}
\lim_{\varepsilon\rightarrow0}\left.  \frac{\partial A}{\partial x}\right\vert
_{-\varepsilon}^{{}}=\lim_{\varepsilon\rightarrow0}\left(  -ik\alpha
\exp(-ik\varepsilon)\right)  =-ik\alpha\text{ .}%
\end{equation}
Hence%
\begin{equation}
\lim_{\varepsilon\rightarrow0}\left.  \frac{\partial A}{\partial x}\right\vert
_{-\varepsilon}^{+\varepsilon}=+2ik\alpha\text{ }.
\end{equation}
Therefore, the amplitude $\alpha$ of the radiation field $A$ emitted from the
charge current sheet of strength $j_{0}$ generated by an incident GR wave is
given by%
\begin{equation}
\alpha=i\frac{1}{2}\frac{\mu_{0}j_{0}}{k}\text{ .} \label{alpha-due-to-j_0}%
\end{equation}

For a very thin film of thickness $d,$ the delta function $\delta\left(
x\right)  $ is approximately%
\begin{equation}
\delta(x)\approx\frac{1}{d}%
\end{equation}
inside the film and zero outside, since then%
\begin{equation}%
{\displaystyle\int\limits_{-d/2}^{d/2}}
\delta(x)dx=1\text{ ,}%
\end{equation}
which implies that%
\begin{equation}
\alpha=i\frac{1}{2}\frac{Z_{0}d}{\omega}j_{\text{e}}\text{ .}
\label{alpha-due-to-j}%
\end{equation}

As we saw in the previous section, an incident GR wave generates within a
superconducting film not only an $E_{\text{G}}$ field but an internal $E$
field as well. In each case, the tangential component of the field must be
continuous across the superconductor-vacuum interface. Since there is no
incoming $E$ field, this continuity condition requires the appearance of an
outgoing $E$ field, which is to say that the charge supercurrent generated by
the GR wave will cause the film to behave like an antenna and radiate EM
waves. For the same sinusoidal time dependence $\exp(-i\omega t)$ of all
fields and currents, and ignoring spatial dependence, we know that%
\begin{equation}
E=-\frac{\partial}{\partial t}A=i\omega A=i\omega\alpha\text{ .}%
\end{equation}
Inserting (\ref{alpha-due-to-j}) into this expression, we see that the
relationship between the charge supercurrent $j_{\text{e}}$ in the current
sheet and the $E$ field both outside and inside the film will be given by%
\begin{equation}
j_{\text{e}}=nqv=-\frac{2}{Z_{0}d}E\text{ .} \label{current-sheet-driven-by-E}%
\end{equation}
The charge conductivity of the film stemming from its behavior as an EM
antenna in the presence of a GR wave is thus given by%
\begin{equation}
\sigma_{\text{e}}=-\frac{2}{Z_{0}d}\text{ .}%
\end{equation}

Now, it must be possible to re-express this charge conductivity as the real
part of the complex mass conductivity. The justification for this step is that
the EM radiation produced in transduction from the incident GR wave leads to
power loss from the wave that escapes to infinity, never to return. Hence the
transduction effect is a lossy process in the GR wave sector, which is no
different from any other irreversible, ohmic process, and can therefore be
characterized as the real part of the mass conductivity. Multiplying each side
of (\ref{current-sheet-driven-by-E}) by $m/q$ and using the relationship
between $E$ and $E_{\text{G}}$ given earlier in (\ref{E_in_terms_of_E_G}), one
finds that the lossy component of the mass current density $\mathbf{j}%
_{\text{G}}$ arising from the transduction of the incident GR wave into an
EM\ wave is given by%
\begin{equation}
j_{\text{loss,G}}=-\frac{2}{Z_{0}d}\frac{m^{2}}{q^{2}}\frac{1}{\Xi
-1}E_{\text{G}}\text{ .}%
\end{equation}
The real part of the mass conductivity $\sigma_{\text{1,G}}$ of the film due
to the dissipative loss by transduction into the escaping EM radiation is
therefore given by%
\begin{equation}
\sigma_{\text{1,G}}=-\frac{2}{Z_{0}d}\frac{m^{2}}{q^{2}}\frac{1}{\Xi-1}%
=\frac{2}{Z_{\text{G}}d}\text{ ,} \label{sigma_1G}%
\end{equation}
where we have taken advantage of the fact that%
\begin{equation}
\Xi-1=\left(  1-\frac{m^{2}Z_{\text{G}}}{q^{2}Z_{0}}\right)  -1=-\frac
{m^{2}Z_{\text{G}}}{q^{2}Z_{0}}\text{ .}%
\end{equation}
We can now use (\ref{sigma_1G}) in conjunction with the nondissipative
conductivity $\sigma_{2,\text{G}}$ given in (\ref{sigma_G_Ksi_reduced})%
\begin{equation}
\sigma_{\text{2,G}}=-\Xi\frac{nq^{2}}{m\omega}\frac{Z_{0}}{Z_{\text{G}}}
\label{sigma_2G}%
\end{equation}
to determine whether loss into EM radiation will undermine the possibility of
GR-wave reflection. We begin by recalling that the full version of the
GR-reflection formula is given by%
\begin{align}
{\mathcal{R}}_{\text{G}}  &  =\left\{  \left(  1+2\frac{\sigma_{1,\text{G}}%
}{\sigma_{1,\text{G}}^{2}+\sigma_{2,\text{G}}^{2}}\frac{1}{Z_{\text{G}}%
d}\right)  ^{2}\right. \nonumber\\
&  \;\quad\left.  +\left(  2\frac{\sigma_{2,\text{G}}}{\sigma_{1,\text{G}}%
^{2}+\sigma_{2,\text{G}}^{2}}\frac{1}{Z_{\text{G}}d}\right)  ^{2}\right\}
^{-1}\text{ .} \label{reflectivity_GR_full}%
\end{align}
Now let us define the parameter $\Sigma,$ which is the dimensionless ratio of
the squares of the two mass conductivities given in (\ref{sigma_1G}) and
(\ref{sigma_2G})%
\begin{align}
\Sigma &  \equiv\left(  \frac{\sigma_{\text{1,G}}}{\sigma_{\text{2,G}}%
}\right)  ^{2}=\left(  \frac{2m\omega}{\Xi nq^{2}Z_{0}d}\right)
^{2}\label{ratio_of_squared_sigmas}\\
&  =\left(  \frac{2}{\pi}\frac{\omega_{\text{d}}}{\omega_{\text{p}}^{^{\prime
}2}}\omega\right)  ^{2}\text{ ,}\nonumber
\end{align}
where $\omega_{\text{p}}^{^{\prime}}$ is the modified plasma frequency
(\ref{modified plasma frequency in terms of Ksi}) and $\omega_{\text{d}}=\pi
c/d$ is a characteristic frequency associated with the thickness of the film
$d$ (i.e., the resonance frequency for its lowest standing-wave mode). In
general, it will be the case that $\Sigma$ is much less than unity when the
frequency of the incident wave is%
\begin{equation}
\omega\ll\frac{\pi}{2}\frac{\omega_{\text{p}}^{^{\prime}2}}{\omega_{\text{d}}%
}=1.1\times10^{16}\text{ rad s}^{-1}\text{.} \label{omega_for_simplification}%
\end{equation}
The microwave frequencies of interest fall well below this limit, so we can
simplify (\ref{reflectivity_GR_full}) to%
\begin{align}
{\mathcal{R}}_{\text{G}}  &  =\left\{  \left(  1+2\frac{\sigma_{1,\text{G}}%
}{\sigma_{2,\text{G}}^{2}}\frac{1}{Z_{\text{G}}d}\right)  ^{2}\right.
\nonumber\\
&  \;\quad\left.  +\left(  2\frac{1}{\sigma_{2,\text{G}}}\frac{1}{Z_{\text{G}%
}d}\right)  ^{2}\right\}  ^{-1}\text{ .} \label{reflectivity_GR_simplified}%
\end{align}
We can then substitute (\ref{sigma_1G}) and (\ref{sigma_2G}) into
(\ref{reflectivity_GR_simplified}) to obtain%
\begin{equation}
{\mathcal{R}}_{\text{G}}=\{\left(  1+\Sigma\right)  ^{2}+\Sigma\}^{-1}\text{
.}%
\end{equation}
For $\omega=2\pi\times(6$ GHz$)$, we see from (\ref{ratio_of_squared_sigmas})
that
\begin{equation}
\Sigma=1.3\times10^{-11} \label{Sigma-numerical-value}%
\end{equation}
and thus that%
\begin{equation}
{\mathcal{R}}_{\text{G}}\approx\left(  1+3\Sigma\right)  ^{-1}=\left(
1+3.8\times10^{-11}\right)  ^{-1}\text{ ,} \label{final_reflectivity_GR}%
\end{equation}
which implies a reflectivity very close to unity. Thus the dissipation (i.e.,
transduction) of an incident GR wave in the form of outgoing EM radiation will
not interfere with the film's ability to specularly reflect GR waves.

As a check on this conclusion, let us examine the ratio $\eta$ of the power
\emph{lost }in the form of outgoing EM radiation to the power \emph{reflected
}in the form of outgoing GR radiation, using the reasonable assumption that
the film acts as a current source in both sectors. Thus,%
\begin{align}
\eta &  =\frac{\left\langle {\mathcal{P}}_{\text{EM}}\right\rangle
}{\left\langle {\mathcal{P}}_{\text{GR}}\right\rangle }=\frac{\left\langle
I_{\text{e}}^{2}\right\rangle Z_{0}}{\left\langle I_{\text{G}}^{2}%
\right\rangle Z_{\text{G}}}=\frac{\left\langle I_{\text{loss,G}}%
^{2}\right\rangle Z_{\text{G}}}{\left\langle I_{\text{G}}^{2}\right\rangle
Z_{\text{G}}}\\
&  =\frac{\left\langle I_{\text{loss,G}}^{2}\right\rangle }{\left\langle
I_{\text{G}}^{2}\right\rangle }=\frac{\left\langle j_{\text{loss,G}}%
^{2}\right\rangle }{\left\langle j_{\text{G}}^{2}\right\rangle }=\frac
{\sigma_{\text{1,G}}^{2}}{\sigma_{\text{2,G}}^{2}}=\Sigma\text{ .}\nonumber
\end{align}
The value for $\Sigma$ given in (\ref{Sigma-numerical-value}) implies that a
negligible fraction of the power of the incoming GR microwave will be lost
through transduction into an outgoing EM wave. A superconducting film at
temperatures sufficiently near $T=0$ will indeed be a highly reflective mirror
for GR microwaves but a highly inefficient transducer of GR microwaves into EM microwaves.

In the parallel case of EM-wave reflection, we can once again take into
account the possibility of transduction by introducing a real term into the EM
conductivity that corresponds to loss into the GR sector (i.e., into an
outgoing GR wave). The resulting real and imaginary parts of the complex
charge conductivity of the film can then be shown to be%
\begin{equation}
\sigma_{\text{1}}=\frac{2}{Z_{\text{0}}d}\text{ \ \ and \ \ }\sigma_{2}%
=\Xi\frac{nq^{2}}{m\omega}\text{ ,} \label{complex-electrical-conductivity}%
\end{equation}
where $\sigma_{\text{1}}$ is the\emph{ }dissipative part of the complex charge
conductivity corresponding to loss by transduction into outgoing GR radiation.
The film will radiate as a GR antenna because of the appearance of a
quadrupolar pattern of \emph{mass }supercurrents (when driven by a
TEM$_{\text{11}}$ incident EM\ plane-wave mode) that couples, via the Cooper
pairs' mass $m=2m_{\text{e}}$, to a gravito-electric field $\mathbf{E}%
_{\text{G}}=-\partial\mathbf{h}/\partial t$ (in the radiation gauge) and to a
gravito-magnetic field $\mathbf{B}_{\text{G}}=\nabla\times\mathbf{h}$, where
$\mathbf{h}$ is the gravitational analog of the electromagnetic vector
potential $\mathbf{A}$.

In fact, (\ref{complex-electrical-conductivity}) leads to an expression for
$\Sigma$ identical to the one given in (\ref{ratio_of_squared_sigmas}). The
subsequent analysis then proceeds unaltered, confirming that the model
developed here is also consistent with the prediction that at temperatures
sufficiently near $T=0$ a superconducting film will be a highly reflective
mirror for EM microwaves but a highly inefficient transducer of EM microwaves
into GR microwaves. Importantly, this prediction is consistent with the
experimental results of Glover and Tinkham \cite{Glover-and-Tinkham}.

\section{Conservation of energy in the reflection process}

Having shown that a superconducting film can specularly reflect a GR wave and
that transduction will not substantially impede this behavior, we turn finally
to the question of whether the expressions for $\sigma_{\text{1,G}}$ and
$\sigma_{\text{2,G}}$ given above in (\ref{sigma_1G}) and (\ref{sigma_2G}) are
consistent with the conservation of energy. This basic physical principle
requires that the absorptivity, reflectivity, and transmissivity of the film
sum to unity:%
\begin{gather}
{\mathcal{A}}_{\text{G}}+{\mathcal{R}}_{\text{G}}+{\mathcal{T}}_{\text{G}%
}=1\label{unity_check}\\
={\mathcal{A}}_{\text{G}}{\mathcal{R}}_{\text{G}}{\mathcal{T}}_{\text{G}%
}\times\left(  \frac{1}{{\mathcal{A}}_{\text{G}}{\mathcal{R}}_{\text{G}}%
}+\frac{1}{{\mathcal{A}}_{\text{G}}{\mathcal{T}}_{\text{G}}}+\frac
{1}{{\mathcal{R}}_{\text{G}}{\mathcal{T}}_{\text{G}}}\right)  \text{
.}\nonumber
\end{gather}
From the analysis presented in Sections 3-6, we know that the reciprocal of
the GR transmissivity is given by%
\begin{equation}
\frac{1}{{\mathcal{T}}_{\text{G}}}=\left(  1+\frac{1}{2}\sigma_{1,\text{G}%
}Z_{\text{G}}d\right)  ^{2}+\left(  \frac{1}{2}\sigma_{2,\text{G}}Z_{\text{G}%
}d\right)  ^{2} \label{1/T_G}%
\end{equation}
and that the reciprocal of the GR reflectivity is given by%
\begin{align}
\frac{1}{{\mathcal{R}}_{\text{G}}}  &  =\left(  1+2\frac{\sigma_{1,\text{G}}%
}{\sigma_{1,\text{G}}^{2}+\sigma_{2,\text{G}}^{2}}\frac{1}{Z_{\text{G}}%
d}\right)  ^{2}\label{1/R_G}\\
&  \;\quad+\left(  2\frac{\sigma_{2,\text{G}}}{\sigma_{1,\text{G}}^{2}%
+\sigma_{2,\text{G}}^{2}}\frac{1}{Z_{\text{G}}d}\right)  ^{2}\text{
.}\nonumber
\end{align}

We can determine the reciprocal of the GR absorptivity ${\mathcal{A}%
}_{\text{G}}$ by considering the work done by a gravito-electric field
$\mathbf{E}_{\text{G}}$ to move a mass $m$ by an infinitesimal displacement
$d\mathbf{x}$:%
\begin{equation}
dW=\mathbf{F}\cdot d\mathbf{x}=m\mathbf{E}_{\text{G}}\cdot d\mathbf{x}\text{
.}%
\end{equation}
The rate of work being done, i.e., the instantaneous power $\mathtt{P}%
$\ delivered by the field to the mass, is%
\begin{equation}
\mathtt{P}=\mathbf{F}\cdot\frac{d\mathbf{x}}{dt}=m\mathbf{E}_{\text{G}}%
\cdot\frac{d\mathbf{x}}{dt}\text{ .}%
\end{equation}
Let $n$ be the number density of mass carriers moving with velocity%
\begin{equation}
\mathbf{v}=\frac{d\mathbf{x}}{dt}\text{ ,}%
\end{equation}
so that the mass current density $\mathbf{j}_{\text{G}}$ is%
\begin{equation}
\mathbf{j}_{\text{G}}=nm\mathbf{v}\text{ .}%
\end{equation}
Then the instantaneous power delivered by the field to the mass carriers per
unit volume moving in a small volume $V$ is%
\begin{equation}
\mathcal{P}=\frac{\mathtt{P}}{V}=nm\mathbf{E}_{\text{G}}\cdot\frac
{d\mathbf{x}}{dt}=\mathbf{j}_{\text{G}}\cdot\mathbf{E}_{\text{G}}\text{\ ,}
\label{P=j.E}%
\end{equation}
where $\mathbf{j}_{\text{G}}$ and $\mathbf{E}_{\text{G}}$ are real quantities.
Let us, however, generalize this expression and represent the current and
field by the complex quantities
\begin{subequations}
\label{cosine functions of time}%
\begin{align}
\mathbf{j}_{\text{G}}  &  \mathbf{=}\mathbf{j}_{0,\text{G}}\exp(-i\omega
t)\text{ and }\label{cosine functions of time-current}\\
\mathbf{E}_{\text{G}}  &  \mathbf{=}\mathbf{E}_{0,\text{G}}\exp(-i\omega
t)\text{ .} \label{cosine functions of time-current-field}%
\end{align}
Then%
\end{subequations}
\begin{equation}
\operatorname{Re}\mathbf{j}_{\text{G}}\mathbf{=}\frac{1}{2}\left[
\mathbf{j}_{0,\text{G}}\exp(-i\omega t)+\mathbf{j}_{0,\text{G}}^{\ast}%
\exp(i\omega t)\right]
\end{equation}
and%
\begin{equation}
\operatorname{Re}\mathbf{E}_{\text{G}}\mathbf{=}\frac{1}{2}\left[
\mathbf{E}_{0,\text{G}}\exp(-i\omega t)+\mathbf{E}_{0,\text{G}}^{\ast}%
\exp(i\omega t)\right]  \text{ .}%
\end{equation}
The \textit{real} instantaneous power per unit volume expressed in terms of
this \emph{complex} current and field is given by
\begin{align}
\mathcal{P}  &  =\operatorname{Re}\mathbf{j}_{\text{G}}\cdot\operatorname{Re}%
\mathbf{E}_{\text{G}}\label{P prod}\\
&  \mathbf{=}\frac{1}{2}\left[  \mathbf{j}_{0,\text{G}}\exp(-i\omega
t)+\mathbf{j}_{0,\text{G}}^{\ast}\exp(i\omega t)\right]  \cdot\nonumber\\
&  \;\;\frac{1}{2}\left[  \mathbf{E}_{0,\text{G}}\exp(-i\omega t)+\mathbf{E}%
_{0,\text{G}}^{\ast}\exp(i\omega t)\right]  \text{.}\nonumber
\end{align}
But the time average over one wave-period $T=2\pi/\omega$ of each second
harmonic term in this expression vanishes because%
\begin{subequations}
\begin{align}
\frac{1}{T}%
{\displaystyle\int\limits_{0}^{T}}
dt\left[  \mathbf{j}_{0,\text{G}}\cdot\mathbf{E}_{0,\text{G}}\exp(-2i\omega
t)\right]   &  =0\label{time-averaged power for real case}\\
\frac{1}{T}%
{\displaystyle\int\limits_{0}^{T}}
dt\left[  \mathbf{j}_{0,\text{G}}^{\ast}\cdot\mathbf{E}_{0,\text{G}}^{\ast
}\exp(+2i\omega t)\right]   &  =0\text{ ,}%
\end{align}
leaving only the DC cross terms%
\end{subequations}
\begin{equation}
\left\langle \mathcal{P}\right\rangle =\frac{1}{4}\left(  \mathbf{j}%
_{0,\text{G}}\cdot\mathbf{E}_{0,\text{G}}^{\ast}+\mathbf{j}_{0,\text{G}}%
^{\ast}\cdot\mathbf{E}_{0,\text{G}}\right)  \text{ ,}%
\end{equation}
which can be re-expressed as%
\begin{equation}
\left\langle \mathcal{P}\right\rangle =\frac{1}{2}\operatorname{Re}%
(\mathbf{j}_{\text{G}}^{\ast}\cdot\mathbf{E}_{\text{G}})\text{ .}
\label{Time-averaged P as Re(j*.E)}%
\end{equation}

Let us apply this result for the time-averaged power density to a
superconducting film by recalling that the relevant gravito-electric field is
the field \textit{inside} the film, so that%
\begin{equation}
\left\langle \mathcal{P}\right\rangle =\frac{1}{2}\operatorname{Re}%
(\mathbf{j}_{\text{G}}^{\ast}\cdot\mathbf{E}_{\text{G-inside}})\text{, }%
\end{equation}
where the gravitational analog of Ohm's law is%
\begin{equation}
\mathbf{j}_{\text{G}}=\sigma_{\text{G}}\mathbf{E}_{\text{G-inside}}\text{ .}%
\end{equation}
Therefore,%
\begin{align}
\left\langle \mathcal{P}\right\rangle  &  =\frac{1}{2}\left(
\operatorname{Re}\sigma_{\text{G}}^{\ast}\right)  (\mathbf{E}_{\text{G-inside}%
}^{\ast}\cdot\mathbf{E}_{\text{G-inside}})\\
&  =\frac{1}{2}\operatorname{Re}\left(  \sigma_{1,\text{G}}-i\sigma
_{2,\text{G}}\right)  \left\vert \mathbf{E}_{\text{G-inside}}\right\vert
^{2}\nonumber\\
&  =\frac{1}{2}\sigma_{1,\text{G}}\left\vert \mathbf{E}_{\text{G-inside}%
}\right\vert ^{2}\text{ .}\nonumber
\end{align}
As in the electromagnetic case discussed in Section 3, the gravito-electric
field inside the film will be related to the incident gravito-electric field
as follows:%
\begin{align}
\mathbf{E}_{\text{G-inside}}  &  =(1-r_{\text{G}})\mathbf{E}%
_{\text{G-incident}}\\
&  =t_{\text{G}}\mathbf{E}_{\text{G-incident}}\nonumber\\
&  =\mathbf{E}_{\text{G-transmitted}}\text{ ,}\nonumber
\end{align}
where $r_{\text{G}}$ is the amplitude reflection coefficient and $t_{\text{G}%
}$ is the amplitude transmission coefficient. Thus the time-averaged power
dissipated inside the entire volume $Ad$ of the film, where $A$ is its area
(an arbitrarily large quantity) and $d$ is its thickness, is given by%
\begin{align}
\left\langle \mathcal{P}\right\rangle Ad  &  =\frac{1}{2}\sigma_{1,\text{G}%
}t_{\text{G}}^{\ast}t_{\text{G}}\left\vert \mathbf{E}_{\text{G-incident}%
}\right\vert ^{2}Ad\\
&  =\frac{A\sigma_{1,\text{G}}d}{2}\mathcal{T}_{\text{G}}\left\vert
\mathbf{E}_{\text{G-incident}}\right\vert ^{2}\text{ ,}\nonumber
\end{align}
where $\mathcal{T}_{\text{G}}=t_{\text{G}}^{\ast}t_{\text{G}}$ is the
transmittivity of the film.

The magnitude of the time-averaged Poynting vector of the incident wave
traveling in the direction $\mathbf{\hat{k}}$ is given by an expression
similar to (\ref{Time-averaged P as Re(j*.E)}), viz.,%
\begin{align}
\left\langle \mathcal{S}\right\rangle  &  =\frac{1}{2}\mathbf{\hat{k}}%
\cdot\operatorname{Re}\left(  \mathbf{E}_{\text{G-incident}}^{\ast}%
\times\mathbf{H}_{\text{G-incident}}\right) \label{Poynting}\\
&  =\frac{1}{2}\frac{1}{Z_{\text{G}}}\operatorname{Re}(\mathbf{E}%
_{\text{G-incident}}^{\ast}\cdot\mathbf{E}_{\text{G-incident}})\nonumber\\
&  =\frac{1}{2Z_{\text{G}}}\left\vert \mathbf{E}_{\text{G-incident}%
}\right\vert ^{2}\text{ ,}\nonumber
\end{align}
from which it follows that the power incident on the area $A$ of the film is%
\begin{equation}
\left\langle S\right\rangle A=\frac{1}{2Z_{\text{G}}}\left\vert \mathbf{E}%
_{\text{G-incident}}\right\vert ^{2}A\text{ .}%
\end{equation}
Thus the absorptivity $\mathcal{A}_{\text{G}}$, which is the ratio of the
time-averaged power dissipated inside the film to the time-averaged power
incident on the film, is given by%
\begin{equation}
\mathcal{A}_{\text{G}}=\frac{\left\langle \mathcal{P}\right\rangle
Ad}{\left\langle S\right\rangle A}=\mathcal{T}_{\text{G}}\sigma_{1,\text{G}%
}Z_{\text{G}}d\text{ .} \label{Absorptivity in terms of transmissivity}%
\end{equation}
Inserting the reciprocal of (\ref{1/T_G}) for $\mathcal{T}_{\text{G}}$ into
(\ref{Absorptivity in terms of transmissivity}) and taking the reciprocal of
the new expression, we find that%
\begin{equation}
\frac{1}{{\mathcal{A}}_{\text{G}}}=\frac{(1+\frac{1}{2}\sigma_{1,\text{G}%
}Z_{\text{G}}d)^{2}+(\frac{1}{2}\sigma_{2,\text{G}}Z_{\text{G}}d)^{2}}%
{\sigma_{1,\text{G}}Z_{\text{G}}d}\text{ .} \label{1/A_G}%
\end{equation}
A calculation confirms that the expressions given in (\ref{1/T_G}),
(\ref{1/R_G}), and (\ref{1/A_G}) are in fact consistent with the requirement
that the absorptivity, reflectivity, and transmissivity sum to unity. When we
insert (\ref{1/T_G}), (\ref{1/R_G}), and (\ref{1/A_G}) into the right-hand
side of (\ref{unity_check}), we obtain a single equation with two variables,
$\sigma_{\text{1,G}}$ and $\sigma_{2,\text{G}}.$ Thus the conservation of
energy in the case of the interaction between a GR wave and a superconducting
film depends solely on the relation between $\sigma_{\text{1,G}}$ and
$\sigma_{2,\text{G}}$.

Recalling the parameter $\Sigma$ introduced in (\ref{ratio_of_squared_sigmas}%
), which characterizes the relation between $\sigma_{\text{1,G}}$ and
$\sigma_{2,\text{G}}$, we can re-express the reciprocals of ${\mathcal{A}%
}_{\text{G}},$ ${\mathcal{R}}_{\text{G}}$, and ${\mathcal{T}}_{\text{G}}$ as
\begin{subequations}
\label{1/ART}%
\begin{align}
\frac{1}{{\mathcal{A}}_{\text{G}}}  &  =2+\frac{1}{2\Sigma}\label{1/A_G-Sigma}%
\\
\frac{1}{{\mathcal{R}}_{\text{G}}}  &  =\frac{2\Sigma^{2}+10\Sigma^{3}%
+8\Sigma^{4}}{2\Sigma^{2}+4\Sigma^{3}+2\Sigma^{4}}\label{1/R_G-Sigma}\\
\frac{1}{{\mathcal{T}}_{\text{G}}}  &  =4+\frac{1}{\Sigma}\text{ .}
\label{1/T_G-Sigma}%
\end{align}
One then finds that%
\end{subequations}
\begin{equation}
{\mathcal{A}}_{\text{G}}{\mathcal{R}}_{\text{G}}{\mathcal{T}}_{\text{G}}%
=\frac{2\Sigma^{2}+4\Sigma^{3}+2\Sigma^{4}}{1+13\Sigma+60\Sigma^{2}%
+112\Sigma^{3}+64\Sigma^{4}} \label{ART}%
\end{equation}
and that%
\begin{gather}
(\frac{1}{{\mathcal{A}}_{\text{G}}{\mathcal{R}}_{\text{G}}}+\frac
{1}{{\mathcal{A}}_{\text{G}}{\mathcal{T}}_{\text{G}}}+\frac{1}{{\mathcal{R}%
}_{\text{G}}{\mathcal{T}}_{\text{G}}})=\label{1/ARTproducts}\\
\text{ \ \ \ \ \ \ }\frac{1+13\Sigma+60\Sigma^{2}+112\Sigma^{3}+64\Sigma^{4}%
}{2\Sigma^{2}+4\Sigma^{3}+2\Sigma^{4}}\text{ .}\nonumber
\end{gather}
But these two expressions are just the reciprocals of one another, confirming
(\ref{unity_check}) in the GR case.

In the EM case, the corresponding real and imaginary parts of the complex
\emph{charge} conductivity are given by%
\begin{align}
\sigma_{1}  &  =\frac{2}{Z_{\text{0}}d}\text{ }\\
\sigma_{2}  &  =\Xi\frac{nq^{2}}{m\omega}\text{ .}%
\end{align}
The fact that the ratio of the squares of these two conductivities is once
again%
\begin{equation}
\left(  \frac{\sigma_{1}}{\sigma_{2}}\right)  ^{2}=\frac{4m^{2}\omega^{2}}%
{\Xi^{2}n^{2}q^{4}Z_{0}^{2}d^{2}}=\Sigma
\end{equation}
is a strong hint that (\ref{unity_check}) will be similarly satisfied in the
EM case. In fact, the expressions given above for $1/\mathcal{A}_{\text{G}},$
$1/\mathcal{R}_{\text{G}},$ and $1/\mathcal{T}_{\text{G}}$ in (\ref{1/ART})
carry over without alteration into the EM case, so that the subsequent steps
of the derivation proceed exactly as above. Consequently, we can say that the
formalism presented here obeys energy conservation both in the case of GR
reflection with EM loss and in the case of EM reflection with GR loss. This is
a strong self-consistency check of the entire calculation.

\section{Experimental and Theoretical Implications}

Most of the experiments presently being conducted on gravitational radiation
aim to passively detect GR waves originating from astrophysical sources. The
specular reflection of GR waves at microwave frequencies from superconducting
thin films due to the Heisenberg-Coulomb effect would allow for a variety of
new experiments, all of which could be performed in a laboratory setting and
some of which would involve mesoscopic quantum objects. Here we identify
several such experiments that should be technologically feasible, commenting
briefly on their interrelations and broader theoretical implications.

Consider first a conceptually simple test of the physics behind the
Heisenberg-Coulomb effect itself. In this experiment, two horizontally
well-separated, noninteracting superconducting bodies are allowed to fall
freely in the non-uniform gravitational field of the Earth. The tidal forces
acting on the two bodies, which are like the tidal forces caused by a
low-frequency GR wave, cause them to converge as they fall freely toward
the center of the earth. Although the gap-protected, global quantum mechanical
phase of the Cooper pairs forces each pair to remain motionless with respect
to the center of mass of its own body, this does nothing to prevent the two
bodies from converging during free fall. The trajectories of these two
superconducting bodies -- recall that they have decohered due to interactions
with their environment and are therefore spatially well separated -- must be
identical to those of \emph{any} two noninteracting, freely falling massive
bodies, in accord with the EP.

Now connect the two bodies by a thin, slack, arbitrarily long, superconducting
wire, so that they become a single, simply-connected, coherent superconducting
system. From a mechanical point of view, the negligible Hooke's constant of
the wire allows each body to move freely, one relative to the other. In this
case, the characteristic frequency of the interaction between the bodies and
the gravitational field, which is given by the inverse of the free-fall time,
is far below the BCS gap frequency and far above the simple harmonic resonance
frequency of the two-bodies-plus-wire system. According to first-order
time-dependent perturbation theory, then, the nonlocalizable Cooper pairs of
the two coherently connected bodies must remain motionless with respect to the
center of mass of the \emph{entire }system, as seen by a distant inertial
observer. This follows, as we have argued in Section 2, from the
gap-protected, global, quantum mechanical phase of the Cooper pairs, which is
at root a consequence of the UP. On the other hand, the ions of the two
coherently connected bodies will attempt to converge toward each other during
free fall, since they want to follow geodesics in accord with the EP.

In this experimental \textquotedblleft tug-of-war\textquotedblright\ between
the Uncertainty Principle and the Equivalence Principle, which principle
prevails? When the temperature is low enough to justify ignoring the effect of
any residual normal electrons (i.e., when the temperature is less than roughly
half the critical temperature, so that the BCS gap is sufficiently close to
its value at absolute zero \cite{Tinkham}), we believe the EP will be
completely overcome by the UP. This must be the case because the charge
separation that would otherwise result as the ions converged while the Cooper
pairs remained motionless (with respect to a distant inertial observer) would
generate an unfavorable, higher-energy configuration of the system. The
\emph{quantum mechanical }Cooper pairs must drag the \emph{classical }ionic
lattice into co-motion with them, so that the coherently connected bodies
\emph{depart }from geodesic motion. That is to say, the bodies must maintain a
\emph{constant distance }from one another as they fall. If two coherently
connected superconducting bodies \emph{were }to converge like any two
noninteracting bodies, one would have to conclude that the UP had failed with
respect to the EP, i.e., that the EP is more universal and fundamental in its
application to all objects than the UP. We do not believe this to be the case.

Theories that propose an \textquotedblleft intrinsic collapse of the
wavefunction\textquotedblright\ or \textquotedblleft objective state
reduction,\textquotedblright\ through some decoherence mechanism, whether by
means of a stochastic process that leverages the entanglement of object and
environment (as originally proposed by Ghirardi, Rimini, and Weber
\cite{GRW}), or by means of a sufficiently large change in the gravitational
self-energy associated with different mass configurations of a system (as
proposed by Penrose \cite{Penrose}), would imply the failure of the
Superposition Principle, and thus of the Uncertainty Principle, in the
experiment outlined above. The existence of any such mechanism would destroy
the Heisenberg-Coulomb effect, but it would also pose a serious problem for
any quantum theory of gravity.

A straightforward geometrical calculation for the free-fall experiment outlined
above shows that the convergence of two noninteracting massive bodies
initially separated by several centimeters would be on the order of microns
for free-fall distances presently attainable in aircraft-based zero-gravity
experiments. Though small, this degree of convergence is readily measureable
by means of laser interferometry. The exact \emph{decrease}, if any, in the
convergence measured for two coherently connected superconducting bodies,
relative to the decrease measured for the same two bodies when the coherent
connection is broken, would allow one to measure the strength of the
Heisenberg-Coulomb effect, with null convergence corresponding to maximal
deflection from free fall.

The specular reflection of GR waves from superconducting films, which we have
argued follows from the Heisenberg-Coulomb effect (see Section 7), might also
allow for the detection of a gravitational Casimir-like force (we thank Dirk
Bouwmeester for this important suggestion). In the EM case, an attractive
force between two nearby metallic plates is created by radiation pressure due
to quantum fluctuations in the EM vacuum energy. If the two plates were made
of a type I superconducting material, it should be possible to detect a change
in the attractive force between them, due to the additional coupling of the
plates to quantum fluctuations in the GR vacuum energy, as the plates were
lowered through their superconducting transition temperature. Observation of
the gravitational analog of the Casimir force could be interpreted as evidence
for the existence of quantum fluctuations in gravitational fields, and hence
as evidence for the need to quantize gravity. If no analog of the Casimir
force were observed despite confirmation of the Heisenberg-Coulomb
effect in free-fall experiments, one would be forced to conclude either that
gravitational fields are not quantizable or that something other than the
Heisenberg-Coulomb effect is wrong with our \textquotedblleft
mirrors\textquotedblright\ argument.

In Sections 8 and 9, we discussed the transduction of GR waves to EM waves and
vice-versa. Although we showed that transduction in either direction will be
highly inefficient in the case of a single superconducting film,
experimentally significant efficiencies in both directions may be attainable
in the case of \emph{a pair of charged superconductors }\cite{Chiao-Townes}.
This would lead to a number of experimental possibilities, all of which employ
the same basic apparatus: two levitated (or suspended) and electrically charged
superconducting bodies that repel one another electrostatically even as they
attract one another gravitationally. For small bodies, it is experimentally
feasible to charge the bodies to \textquotedblleft
criticality,\textquotedblright\ i.e., to the point at which the forces of
repulsion and attraction cancel \cite{Chiao-Townes}. At criticality, the
apparatus should become an effective transducer of incoming GR radiation,
i.e., it should enable 50\% GR-to-EM transduction efficiency. By time-reversal
symmetry, it should also become an effective transducer of incoming EM
radiation, i.e., it should also enable 50\% EM-to-GR transduction efficiency.
Chiao has previously labeled this type of apparatus a \textquotedblleft
quantum transducer\textquotedblright\ \cite{Chiao-Townes}.

Two variations on a single-transducer experiment could provide new and
compelling, though still indirect, evidence for the existence of GR waves.
First, an electromagnetically isolated transducer should generate an EM signal
in the presence of an incoming GR wave, since the transducer should convert
half the power contained in any incoming GR wave into a detectable outgoing EM
wave. This might allow for the detection of the cosmic gravitational-wave
background (CGB) at microwave frequencies, assuming that certain cosmological
models of the extremely early Big Bang are correct \cite{Chiao-NASA}. If no
transduced EM signal were detected despite confirmation of the H-C effect, one
would be forced to conclude either that something is wrong with the
\textquotedblleft mirrors\textquotedblright\ argument or the GR-to-EM
\textquotedblleft transduction\textquotedblright\ argument, or that there is
no appreciable CGB at the frequency of investigation.

A single quantum transducer should also behave anomalously below its
superconducting transition temperature in the presence of an incoming EM wave
(we thank Ken Tatebe for this important suggestion). By the principle of the
conservation of energy, an EM receiver directed at the transducer should
register a significant \emph{drop }in reflected power when the transducer is
\textquotedblleft turned on\textquotedblright\ by lowering its temperature
below the transition temperature of the material, since energy would then be
escaping from the system in the form of invisible (transduced) GR waves. If no
drop in reflected power were observed despite confirmation of GR-to-EM
transduction in the experiment outlined in the previous paragraph, one would
need to reconsider the validity of the principle of time-reversal symmetry in
the argument for EM-to-GR transduction.

Finally, if an efficient quantum transducer were to prove experimentally
feasible, two transducers operating in tandem would open up the possibility of
GR-wave communication. As a start, a gravitational Hertz-like experiment
should be possible. An initial transducer could be used to partially convert
an incoming EM into an outgoing GR wave. A second transducer, spatially
separated and electromagnetically isolated from the first, could then be used
to partially back-convert the GR wave generated by the first transducer into a
detectable EM wave. The same two-transducer arrangement could also be used to
confirm the predicted speed and polarization of GR waves. Of course, wireless
communication via GR waves would be highly desirable, since all normal matter
is effectively transparent to GR radiation. Such technology would also open up
the possibility of wireless power transfer over long distances. On the other
hand, if a Hertz-like arrangement were to yield a null result despite the
success of the previously outlined single-transducer experiments, one would
infer that the success of those experiments was due to something other than
the existence of GR waves.

In summary, a new class of laboratory-scale experiments at the interface of
quantum mechanics and gravity follows if the argument presented here for
superconducting GR-wave mirrors is correct. Such experiments could be a boon
to fundamental physics. For example, one could infer from the experimental
confirmation of a gravitational Casimir effect that gravitational fields are
in fact quantized. Confirmation of the Heisenberg-Coulomb effect would also
point to the need for a unified \emph{gravito-electrodynamical} theory for
weak, but quantized, gravitational and electromagnetic fields interacting with
nonrelativistic quantum mechanical matter. Such a theory would fall far short
of the ultimate goal of unifying all known forces of nature into a
\textquotedblleft theory of everything,\textquotedblright\ but it would
nonetheless be a very useful theory to have.

\appendix

\section{The magnetic and kinetic inductances of a thin metallic film}

The EM inductance $L$ of the superconducting film is composed of two parts:
the magnetic inductance $L_{\text{m}}$, which arises from the magnetic fields
established by the charge supercurrents, carried by the Cooper pairs, and the
kinetic inductance $L_{\text{k}}$, which arises from the Cooper pairs'
inertial mass \cite{Merservey-and-Tedrow}. Using (\ref{l_k}) and the values of
$\xi_{0}=83$ nm and $\delta_{\text{p}}=\lambda_{\text{L}}=37$ nm for Pb at
microwave frequencies, one finds for our superconducting film that
$l_{\text{k}}$ is on the order of $10^{-5}$ m and that $L_{\text{k}}$ is on
the order of $10^{-11}$ henries.

$L_{\text{m}}$ can be found using the magnetic potential energy relations%
\begin{equation}
U=\int\frac{B^{2}}{2\mu_{0}}\text{ }d^{3}x=\frac{1}{2}L_{\text{m}}I^{2}\text{
},
\end{equation}
where $U$ is the magnetic potential energy, $B$ is the magnetic induction
field, and $I$ is the (uniform) current flowing through the film. Thus,
\begin{equation}
L_{\text{m}}=\int\frac{B^{2\text{ }}}{I^{2}\mu_{0}}d^{3}x\text{ }.
\end{equation}
A closed-form, symbolic expression for this integral is complicated for the
geometry of a film, but numerical integration shows that in the case of a Pb
film with dimensions $1$ cm $\times$ $1$ cm $\times$ $2$ nm, $L_{\text{m}}$ is
on the order of at most $10^{-15}$ henries, which is much smaller than
$L_{\text{k}}$. The experiments of Glover and Tinkham
\cite{Glover-and-Tinkham} corroborate the validity of this approximation.
Thus, we can safely neglect the magnetic inductance $L_{\text{m}}$ in our
consideration of $L$.

A comparison of this result for $L_{\text{m}}$ with the result for
$L_{\text{m,G}}$ in the gravitational sector reveals that%
\begin{equation}
\frac{L_{\text{m,G}}}{L_{\text{m}}}=\frac{\mu_{\text{G}}}{\mu_{0}}\ \text{.}
\label{ratio of mag-inds}%
\end{equation}
Recall now that the expression for $l_{\text{k,G}}^{^{\prime}}$ given by
(\ref{l_kG-plasma-corr-factor}) is%
\begin{equation}
l_{\text{k,G}}^{^{\prime}}=d\left(  \frac{\delta_{\text{p}}}{d}\right)
^{2}\text{ ,}%
\end{equation}
which is just the expression for $l_{\text{k,p}}$ ($\approx l_{\text{k}}$)
derived in Appendix B below. Thus we see that%
\begin{equation}
\frac{L_{\text{k,G}}}{L_{\text{k}}}=\frac{\mu_{\text{G}}l_{\text{k,G}%
}^{^{\prime}}}{\mu_{0}l_{\text{k}}}\approx\frac{\mu_{\text{G}}}{\mu_{0}%
}\ \text{.} \label{ratio of kin-inds}%
\end{equation}
From (\ref{ratio of mag-inds}) and (\ref{ratio of kin-inds}), it follows that%
\begin{equation}
\frac{L_{\text{m,G}}}{L_{\text{k,G}}}\approx\frac{L_{\text{m}}}{L_{\text{k}}%
}\text{ .}%
\end{equation}
Thus, we can also safely neglect the gravito-magnetic inductance
$L_{\text{m,G}}$ in our consideration of $L_{\text{G}}.$

\section{The kinetic inductance length scale in a collisionless plasma model}

In this appendix we ignore the quantum mechanical properties of
superconducting films and consider the simpler, classical problem of the
kinetic inductance (per square) of a thin metallic film. We begin with a
physically intuitive derivation of the kinetic inductance length scale
$l_{\text{k}}$ due to D. Scalapino (whom we thank for pointing out this
derivation to us). The current density for a thin metallic film is given by%
\begin{equation}
j=n_{\text{e}}ev=\frac{I}{A}=\frac{I}{wd}\text{ ,}%
\end{equation}
where $e$ is the electron charge, $v$ is the average velocity of the
electrons, $n_{\text{e}}$ is the number density, $A$ is the cross-sectional
area of the film through which the current flows, $w$ is film's width, and $d$
is its thickness. The velocity of the electrons within the film can then be
expressed as%
\begin{equation}
v=\frac{I}{w}\frac{1}{n_{\text{e}}ed}=\frac{I_{\text{w}}}{n_{\text{e}}%
ed}\text{ ,} \label{electron-vel-thin-film}%
\end{equation}
where $I_{w}$ is the current per width. Now, by conservation of energy it must
be the case that%
\begin{equation}
\frac{L_{\text{k}}I_{\text{w}}^{2}}{2}=\frac{m_{\text{e}}v^{2}}{2}n_{\text{e}%
}d\text{ .} \label{kin-induct-energy-cons}%
\end{equation}
The left-hand side of (\ref{kin-induct-energy-cons}) gives the energy per
square meter carried by the film's electrons in terms of the film's kinetic
inductance per square and the square of the current per width, whereas the
right-hand side gives the same quantity in terms of the kinetic energy per
electron multiplied by the number of electrons per square meter of the film.
Substituting (\ref{electron-vel-thin-film}) into (\ref{kin-induct-energy-cons}%
) and recalling the expression for the plasma skin depth given in
(\ref{plas-skin-dep}), one finds that%
\begin{equation}
L_{\text{k}}=\frac{m_{\text{e}}}{n_{\text{e}}e^{2}d}=\mu_{0}\frac
{\delta_{\text{p}}^{2}}{d}\text{ ,}%
\end{equation}
which implies that the kinetic inductance length scale of the film is given by%
\begin{equation}
l_{\text{k}}=\frac{L_{\text{k}}}{\mu_{0}}=d\left(  \frac{\delta_{\text{p}}}%
{d}\right)  ^{2}\text{ .} \label{kin-ind-length-scale}%
\end{equation}

Now let us derive the kinetic inductance length scale of a thin
superconducting film by treating the film as though it were a neutral,
collisionless plasma consisting of Cooper-paired electrons moving
dissipationlessly through a background of a positive ionic lattice. We assume
that the film is at absolute zero temperature and that the mass of each
nucleus in the lattice is so heavy that, to a good first approximation, the
motion of the lattice in response to an incident EM wave can be neglected when
compared to the motion of the electrons. If one then analyzes the film's
response to the incident EM wave using the concepts of polarization and
susceptibility, it is possible to show for all non-zero frequencies that%
\begin{equation}
\sigma_{1}=0\text{ \ \ and \ \ }\sigma_{2}=\varepsilon_{0}\frac{\omega
_{\text{p}}^{2}}{\omega}\text{ .}%
\end{equation}
Recalling the basic relationship between the kinetic inductance $L_{\text{k}}$
and $\sigma_{2}$ given in (\ref{inductive-reactance}), as well as the fact
that $\mu_{0}=1/\varepsilon_{0}c^{2}$, and that $\delta_{\text{p}}%
=c/\omega_{\text{p}}$ when $\omega\ll\omega_{\text{p}}$, we see that according
to this model the kinetic inductance of the superconducting film (in the limit
of $\omega\ll\omega_{\text{p}}$) $L_{\text{k,p}}$ is given by%
\begin{equation}
L_{\text{k,p}}=\frac{1}{\varepsilon_{0}\omega_{\text{p}}^{2}d}=\mu_{0}%
\frac{\delta_{\text{p}}^{2}}{d}\text{ ,}%
\end{equation}
which implies that the plasma version of the kinetic inductance length scale
$l_{\text{k,p}}$ for a superconducting film at absolute zero is%
\begin{equation}
l_{\text{k,p}}=\frac{L_{\text{k,p}}}{\mu_{0}}=d\left(  \frac{\delta_{\text{p}%
}}{d}\right)  ^{2} \label{l_k-plasma}%
\end{equation}
in agreement with (\ref{kin-ind-length-scale}). The discrepancy between these
expressions and the one obtained in (\ref{l_k}) in Section 4 on the basis of
the more sophisticated BCS model,%
\begin{equation}
l_{\text{k}}=\xi_{0}\left(  \frac{\delta_{\text{p}}}{d}\right)  ^{2}\text{ ,}
\label{l_k-appendix-B}%
\end{equation}
arises from the fact that the classical approaches taken here know nothing of
the additional length scale of the BCS theory, namely, the coherence length
$\xi_{0}.$ This quantum mechanical length scale is related to the BCS energy
gap $\Delta$ through (\ref{coh-len}) and cannot enter into derivations based
solely on classical concepts; hence the appearance of the prefactor $d$
instead of $\xi_{0}$ in (\ref{kin-ind-length-scale}) and (\ref{l_k-plasma}).

\section{Impedance and scattering cross-section}

The relevance of the concept of impedance to the question of scattering
cross-section can be clarified by considering the case of an EM plane wave
scattered by a Lorentz oscillator, which plays a role analogous to the
resonant bar in Weinberg's considerations of GR-wave scattering
\cite{Weinberg}. The Poynting vector $\mathbf{S}$ of the incident EM wave is
related to the impedance of free space $Z_{0}$ as follows:%
\begin{equation}
\mathbf{S=E\times H=}\frac{1}{Z_{0}}E^{2}\mathbf{\hat{k}}\text{ ,}
\label{poynting-vector}%
\end{equation}
where the wave's electric field $\mathbf{E}$ and magnetic field $\mathbf{H}$
are related to one another by $\left\vert \mathbf{E}\right\vert \mathbf{=}%
Z_{0}\left\vert \mathbf{H}\right\vert $, where $Z_{0}=\sqrt{\mu_{0}%
/\varepsilon_{0}}=$ 377 ohms is the characteristic impedance of free space,
and where $\mathbf{\hat{k}}$\ is the unit vector denoting the direction of the
wave's propagation.

Multiplying the scattering cross-section $\sigma$ (not to be confused with the
conductivity) by the time-averaged magnitude of the Poynting vector
$\left\langle S\right\rangle $, which is the average energy flux of the
incident wave, we get the time-averaged power $\left\langle P\right\rangle $
scattered by the oscillator, viz.,%
\begin{equation}
\sigma\left\langle S\right\rangle =\left\langle P\right\rangle =\sigma
\left\langle E^{2}\right\rangle /Z_{0}\text{ },
\end{equation}
where the angular brackets denote a time average over one cycle of the
oscillator. It follows that%
\begin{equation}
\sigma=\frac{Z_{0}\left\langle P\right\rangle }{\left\langle E^{2}%
\right\rangle }\text{ .}%
\end{equation}
When driven on resonance, a Lorentz oscillator dissipates an amount of power
given by%
\begin{equation}
\left\langle P\right\rangle =\left\langle eE\frac{dx}{dt}\right\rangle
=\frac{\left\langle E^{2}\right\rangle }{\gamma m_{\text{e}}/e^{2}}\text{ },
\end{equation}
where $x$ denotes the oscillator's displacement, $e$ is the charge of the
electron, $m_{\text{e}}$ is its mass, and $\gamma$ is the oscillator's
dissipation rate. The oscillator's EM scattering cross-section is thus related
to $Z_{0}$ as follows:%
\begin{equation}
\sigma=\frac{Z_{0}}{\gamma m_{\text{e}}/e^{2}}\text{ }.
\label{cross-section-and-impedance}%
\end{equation}

Maximal scattering will occur when the dissipation rate of the oscillator
$\gamma$ and thus $\gamma m_{\text{e}}/e^{2}$ are minimized. In general, one
can minimize the dissipation rate of an oscillator by minimizing its ohmic or
dissipative resistance, which is a form of impedance. Hence Weinberg suggested
using dissipationless superfluids instead of aluminum for the resonant bar,
and we suggest here using zero-resistance superconductors instead of
superfluids. In particular, Weinberg's analysis showed that if the damping of
the oscillator is sufficiently dissipationless, such that radiation damping by
GR radiation becomes dominant, the cross-section of the oscillator on
resonance is on the order of a square wavelength, and is independent of
Newton's constant $G$. However, the bandwidth of the resonance is extremely
narrow, and is directly proportional to $G$.

In this regard, an important difference between neutral superfluids and
superconductors is the fact that the electrical charge of the Cooper pairs
enters into the interaction of the superconductor with the incoming GR wave.
This leads to an enormous enhancement of the oscillator strength of Weinberg's
scattering cross-section extended to the case of a superconductor in its
response to the GR wave, relative to that of a neutral superfluid or of normal
matter like that of a Weber bar.

As we have seen earlier, the non-localizability of the negatively charged
Cooper pairs, which follows from the Uncertainty Principle and is protected by
the BCS energy gap, causes them to undergo \emph{non-geodesic} motion in
contrast to the decoherence-induced \emph{geodesic }motion of the positively
charged ions in the lattice, which follows from the Equivalence Principle. The
resulting charge separation leads to a virtual plasma excitation inside the
superconductor. The enormous enhancement of the conductivity that follows from
this, i.e., the H-C effect, can also be seen from the infinite-frequency sum
rule that follows from the Kramers-Kronig relations, which are based on
causality and the linearity of the response of the superconductor to either an
EM or a GR wave \cite[p. 88, first equation]{Tinkham}.

In the electromagnetic sector, the Kramers-Kronig relations for the real part
of the charge conductivity $\sigma_{1}(\omega)$ and the imaginary part
$\sigma_{2}(\omega)$ (not to be confused with the above scattering
cross-section $\sigma$) are given by \cite[p. 279]{LxL}
\begin{subequations}
\label{KK-relations}%
\begin{align}
\sigma_{1}(\omega)  &  =\frac{2}{\pi}%
{\displaystyle\int\limits_{0}^{\infty}}
\frac{\omega^{\prime}\sigma_{2}\left(  \omega^{\prime}\right)  d\omega
^{\prime}}{\omega^{\prime2}-\omega^{2}}\label{1st KK relation from sigma_1}\\
\sigma_{2}(\omega)  &  =-\frac{2\omega}{\pi}%
{\displaystyle\int\limits_{0}^{\infty}}
\frac{\sigma_{1}\left(  \omega^{\prime}\right)  d\omega^{\prime}}%
{\omega^{\prime2}-\omega^{2}}\text{ .} \label{2nd KK relation for sigma_2}%
\end{align}
From (\ref{2nd KK relation for sigma_2}) and the fact that electrons become
free particles at infinitely high frequencies, one can derive the
infinite-frequency sum rule given by Kubo \cite{LxL,Kubo}
\end{subequations}
\begin{equation}%
{\displaystyle\int\limits_{0}^{\infty}}
\sigma_{1}(\omega)d\omega=\frac{\pi}{2}\varepsilon_{0}\omega_{\text{p}}%
^{2}\text{ , where }\omega_{\text{p}}^{2}=\frac{n_{\text{e}}e^{2}}%
{\varepsilon_{0}m_{\text{e}}}\text{ .} \label{Kubo sum rule}%
\end{equation}

In the GR sector, making the replacement in (\ref{Kubo sum rule}),%
\begin{equation}
\frac{e^{2}}{4\pi\varepsilon_{0}}\rightarrow Gm^{2}\text{ ,}
\label{sum-rule-replacement}%
\end{equation}
where $m$ is regarded as the mass of the neutral atom that transports the mass
current within the superfluid, is relevant to the interaction between a
neutral superfluid and an incident GR wave. This leads to the following
infinite-frequency sum rule:%
\begin{equation}%
{\displaystyle\int\limits_{0}^{\infty}}
\sigma_{1,\text{G}}(\omega)d\omega=2\pi^{2}n\varepsilon_{0}Gm\text{ .}%
\end{equation}
Numerically, this result is extremely small relative to the result given in
(\ref{Kubo sum rule}), which implies a much narrower scattering cross-section
bandwidth in the GR sector.

In the case of a superconductor, the replacement given by
(\ref{sum-rule-replacement}) is unphysical, due to the charged nature of its
mass carriers, i.e., Cooper pairs. Here, Kramers-Kronig relations similar to
those given in (\ref{KK-relations}) lead to a result identical to the one
given in (\ref{Kubo sum rule}). Thus, using superconductors in GR-wave
detectors will lead to bandwidths of scattering cross-sections that are orders
of magnitude broader than those of neutral superfluids.

One important implication of this argument concerns the GR scattering
cross-section of a superconducting sphere. If the sphere's circumference is on
the order of a wavelength of an incident GR wave, the wave will undergo the
first resonance of Mie scattering. In the case of specular reflection from the
surface of a superconducting sphere, this corresponds to a broadband,
geometric-sized scattering cross-section, i.e., a scattering cross-section on
the order of a square wavelength over a wide bandwidth. This implies that two
charged, levitated superconducting spheres in static mechanical equilibrium,
such that their electrostatic repulsion balances their gravitational
attraction, should become an efficient transducer for converting EM\ waves
into GR waves and vice versa \cite{Chiao-Townes}. As suggested in Section 10,
two such transducers could be used to perform a Hertz-like experiment for GR microwaves.

\section*{Acknowledgements}

We thank Jim Bardeen, Dirk Bouwmeester, Amir Caldeira, Sang Woo Chi, Victoria
Chiu, Spencer De Santo, Ivan Deutsch, Uwe Fischer, Theodore Geballe, Vesselin
Gueorguiev, Natalie Hall, Jim Hartle, Gary Horowitz, Boaz Ilan, Joseph Imry,
Derrick Kiley, Hagen Kleinert, Don Marolf, Luis Martinez, Kevin Mitchell,
Giovanni Modanese, Sir Roger Penrose, Clive Rowe, Doug Scalapino, Nils
Schopohl, Achilles Speliotopoulos, Gary Stephenson, and Ken Tatebe for their
many helpful comments and criticisms. R.Y.C. thanks Peter Keefe, Theo
Nieuwenhuizen, and Vaclav Spicka for the invitation to speak at the
\emph{Frontiers of Quantum and Mesoscopic Thermodynamics '08} conference in
Prague on the subject of this paper. This work was supported in part by a
STARS Planning Grant and a STARS Research Grant from the Center for Theology
and the Natural Sciences.


\begin{thebibliography}{99}
\bibitem {wrong-interp-of-EP}An interpretation of \textquotedblleft the
universal validity of the EP\textquotedblright\ that ignores the existence of
forces other than gravity, and also ignores quantum mechanics, would lead to a
universal length contraction or expansion between all parts of a piece of
matter in response to an incident GR wave. This would nullify the non-geodesic
motion of the Cooper pairs, but it would also nullify all forms of rigidity,
including those of an ordinary ruler. This interpretation of the EP is clearly
false, as demonstrated by the fact that the expansion of the Universe in the
Friedmann metric would accordingly become meaningless.
\bibitem {delocalization}It should be noted that a superconductor's normal
conduction electrons are just as localizable within the material as the ions,
since no energy gap exists in their case. The behavior of normal conduction
electrons is said by condensed-matter physicists to be \textquotedblleft
delocalized\textquotedblright\ with respect to a given ion within a few
lattice constants, but these electrons are not \textquotedblleft
nonlocalizable\textquotedblright\ in the sense of the Cooper pairs discussed here.
\bibitem {conductivity-sign-convention}Tinkham \cite{Tinkham} uses a different
convention, namely, $\sigma_{s}(\omega)=\sigma_{1\text{s}}(\omega
)-i\sigma_{2\text{s}}(\omega)$, which arises from his choice for time
dependence to be $\exp\left(  +i\omega t\right)  $, whereas we employ a time
dependence of $\exp\left(  -i\omega t\right)  $, so that here $\sigma
_{\text{s}}(\omega)=\sigma_{1\text{s}}(\omega)+i\sigma_{2\text{s}}(\omega).$
\bibitem {Sign-of-mass-current}The electron's $negative$ charge means that the
charge current density $\mathbf{j}_{\text{e}}$ flows in a direction $opposite$
to the velocity vector $\mathbf{v}$ of the electrons, whereas the electron's
$positive$ mass means that the mass-current density $\mathbf{j}_{\text{G}}$
flows in the $same$ direction as $\mathbf{v}$. Hence the mass-current density
$\mathbf{j}_{\text{G}}$ flows in a direction $opposite$ to the charge current
density $\mathbf{j}_{\text{e}}$. But because of the negative sign in front of
the mass current density source term on the right-hand side of the
gravitational Ampere-like law in the Maxwell-like equations
(\ref{Maxwell-like-eq-4}), $\mathbf{\nabla\times B}_{\text{G}}=-\mu_{\text{G}%
}\mathbf{j}_{\text{G}}$, there is a minus sign on the right-hand side of
(\ref{Ampere's-law}) in the GR case, which cancels the former minus sign, so
that on the right-hand side of (\ref{2nd-r-t-relation}) and in all subsequent
equations for the GR case, the signs remain unchanged. The minus sign in front
of the mass-current-density source term on the right-hand side of the
Ampere-like law (\ref{Maxwell-like-eq-4}) in the Maxwell-like equations stems
from the fact that, when one takes the divergence of both sides of this
equation, one must obtain zero, since the divergence of a curl is zero. This
will indeed be the result, provided that both the continuity equation for the
mass current density and the Gauss-like law (\ref{Maxwell-like-eq-1}) in the
Maxwell-like equations are simultaneously satisfied on the right-hand side of
the resulting equation. The minus sign in front of the mass-density source
term in the Gauss-like law is necessitated by the experimental fact that the
gravitational force between all masses is attractive, which implies that the
minus sign in front of the mass-current-density source term in the Ampere-like
law (\ref{Maxwell-like-eq-4}) is also necessitated by the same experimental
fact. These two minus signs also emerge naturally from the linearization of
Einstein's field equations.
\bibitem {transmission-line-reflectivity}This formula can also be understood
in terms of a simple transmission line model, where the line has a
characteristic impedance $Z$ and a purely reactive shunt element $iX$, where
the element $iX$ could represent the effect of the kinetic inductance of the
film. We wish to calculate the power reflected from this element when an
incident wave propagates down the transmission line. The equivalent circuit
consists of the same transmission line terminated by means of a parallel
combination of the reactive element and a resistor having a resistance $Z$,
which is a real number. The reflectivity of this configuration is given by%
\[
{\mathcal{R}}=\left\vert \rho\right\vert ^{2}=\rho^{\ast}\rho=\left\vert
\frac{Z-Z_{\text{eff}}}{Z+Z_{\text{eff}}}\right\vert ^{2}%
\]
where%
\[
\rho=\frac{Z-Z_{\text{eff}}}{Z+Z_{\text{eff}}}%
\]
and%
\[
Z_{\text{eff}}=\frac{iX\cdot Z}{iX+Z}\text{ .}%
\]
From this it follows that%
\begin{align*}
\rho &  =\left(  Z-\frac{iX\cdot Z}{iX+Z}\right)  \left(  Z+\frac{iX\cdot
Z}{iX+Z}\right)  ^{-1}\\
&  =\frac{1}{1+2iX/Z}%
\end{align*}
and that%
\begin{align}
{\mathcal{R}}  &  =\left(  \frac{1}{1-2iX/Z}\right)  \left(  \frac{1}%
{1+2iX/Z}\right) \nonumber\\
&  =\left\{  1+\left(  2\frac{X}{Z}\right)  ^{2}\right\}  ^{-1}\text{
.}\nonumber
\end{align}
With the identifications of $X=X_{\text{L}}$ and $Z=Z_{\text{0}}$, we see that
the reflectivity of the reactive element has the same form as the EM
reflectivity of the superconducting film given by
(\ref{EM-reflectivity-imped-ratio}).
\bibitem {Drude Model}The Drude model involves an equation of motion for the
expectation value of the electron velocity, namely%
\[
m_{\text{e}}\frac{d}{dt}\left\langle \mathbf{v}_{\text{e}}\right\rangle
=\frac{-m_{\text{e}}}{\tau}\left\langle \mathbf{v}_{\text{e}}\right\rangle
-e\mathbf{E}\text{ ,}%
\]
where one adopts a semi-classical approach in which the electrons are treated
quantum-mechanically and the external fields classically. This Langevin-like
equation follows from the Fokker-Planck equation, which models diffusion in
phase-space. The mean free time $\tau$ in the Drude model is assumed to be
$d/v_{\text{F}}$ where $d$ is the thickness of the film and $v_{\text{F}}$ is
the Fermi velocity. The steady-state solution of this model is given by%
\[
\left\langle \mathbf{v}_{\text{e}}(t\rightarrow\infty)\right\rangle
=\frac{\tau e}{m_{\text{e}}}\mathbf{E}\text{ }\mathbf{.}%
\]
If we now use the usual expression for the current density%
\[
\left\langle \mathbf{j}_{\text{e}}\right\rangle =n_{\text{e}}e\left\langle
\mathbf{v}_{\text{e}}\right\rangle
\]
where $n_{\text{e}}$ is the number density of the electrons, we see that%
\[
\left\langle \mathbf{j}_{\text{e}}\right\rangle =\frac{n_{\text{e}}e^{2}%
d}{m_{\text{e}}v_{\text{F}}}\mathbf{E}\text{ }\mathbf{.}%
\]
Equating $\sigma_{\text{n}}$ with the prefactor on the right-hand side, we
recover Ohm's law and arrive at the expression given in
(\ref{normal-conductivity}) for the normal conductivity.
\bibitem {Schiff-Barnhill}Note that the $\mathbf{j}_{\text{G}}$ and
$\mathbf{F}_{\text{tot}}$ will be nonzero as long as the Schiff-Barnhill
effect \cite{DeWitt}%
\[
q\mathbf{E}+m\mathbf{g=0}\text{ ,}%
\]
which is present in a normal metal, is absent in the superconductor. The
static mechanical equilibrium represented by the equation above is never
achieved inside a superconductor, since the establishment of such equilibrium
requires the action of dissipative processes that are present in a normal
metal but absent in a superconductor. In particular, we should expect a large
breakdown of the Schiff-Barnhill effect at microwave frequencies, when plasma
oscillations of the superconductor can in fact be virtually excited purely electromagnetically.
\bibitem {density-operator-EOM}One should regard
(\ref{EOM for rho_e and rho_G}) as a quantum-mechanical operator equation of
motion for the Cooper pair's charge density operator $\rho_{\text{e}}$\ and
mass density $\rho_{\text{G}}$\ operator, which are related to the
quantum-mechanical density operator $\rho=\left\vert \psi\right\rangle
\left\langle \psi\right\vert $ by%
\[
\rho_{\text{e}}=q\left\vert \psi\right\rangle \left\langle \psi\right\vert
\text{ and }\rho_{\text{G}}=m\left\vert \psi\right\rangle \left\langle
\psi\right\vert \text{ ,}%
\]
respectively. Similarly, the probability current density $\mathbf{j}$, the
quantum velocity field $\mathbf{v}$, and the acceleration field $\mathbf{a}$
of the superconductor are also quantum operators.
\bibitem {sigma at DC}This applies to non-zero frequencies only, since at DC,
the mass conductivity $\sigma_{1\text{,G}}$, like the charge conductivity, is
infinite, i.e., where
\[
\sigma_{1\text{,G}}=A_{\text{G}}\delta\left(  \omega\right)
\]
for some constant $A_{\text{G}}$.
\end{thebibliography}
\end{document}